\documentclass[twocolumn,usenames,dvipsnames]{aastex631}

\usepackage{longtable}
\usepackage{amsmath}
\usepackage[autostyle, english = american]{csquotes}
\usepackage[figuresleft]{rotating}

\MakeOuterQuote{"}

\shorttitle{APO-K2 Catalog}
\shortauthors{Schonhut-Stasik et al.}
\graphicspath{{./}{figures/}}

\begin{document}

\title{The APO-K2 Catalog. I. 7,673 Red Giants with Fundamental Stellar Parameters from APOGEE DR17 Spectroscopy and K2-GAP Asteroseismology}

\correspondingauthor{Jessica Schonhut-Stasik}
\email{jessica.s.stasik@vanderbilt.edu}

\author[0000-0002-1043-8853]{Jessica Schonhut-Stasik}
\altaffiliation{Neurodiversity Inspired Science and Engineering Graduate Fellow.}
\affiliation{Vanderbilt University, Department of Physics \& Astronomy, 6301 Stevenson Center Ln., Nashville, TN 37235, USA} 

\author[0000-0002-7550-7151 is my ORC-ID!]{Joel C. Zinn}
\altaffiliation{NSF Astronomy and Astrophysics Postdoctoral Fellow.}
\affiliation{Department of Astrophysics, American Museum of Natural History, Central Park West at 79th Street, New York, NY 10024, USA}

\author[0000-0002-3481-9052]{Keivan G. Stassun}
\affiliation{Vanderbilt University, Department of Physics \& Astronomy, 6301 Stevenson Center Ln., Nashville, TN 37235, USA}

\author{Marc Pinsonneault}
\affiliation{Department of Astronomy, The Ohio State University, McPherson Laboratory, 140 W 18th Ave, Columbus, OH 43210, USA}

\author{Jennifer A. Johnson}
\affiliation{Department of Astronomy, The Ohio State University, McPherson Laboratory, 140 W 18th Ave, Columbus, OH 43210, USA}
\affiliation{Center for Cosmology and AstroParticle Physics, Physics Research Building, 190 West Woodruff Ave, Columbus, OH 43210, USA}

\author[0000-0003-1634-4644]{Jack T. Warfield}
\affiliation{Department of Astronomy, The University of Virginia, 530 McCormick Road, Charlottesville, VA 22904, USA}

\author{Dennis Stello}
\affiliation{School of Physics, University of New South Wales, NSW 2052, Australia}
\affiliation{Sydney Institute for Astronomy (SIfA), School of Physics, University of Sydney, NSW 2006, Australia}
\affiliation{Stellar Astrophysics Centre, Department of Physics and Astronomy, Aarhus University, DK-8000 Aarhus C, Denmark}

\author{Yvonne Elsworth}
\affiliation{Stellar Astrophysics Centre, Department of Physics and Astronomy, Aarhus University, Ny Munkegade 120, DK-8000 Aarhus C,
Denmark}
\affiliation{School of Physics and Astronomy, University of Birmingham, Edgbaston, Birmingham, B15 2TT, UK}

\author[0000-0002-8854-3776]{Rafael A. Garc\'{i}a}
\affil{Universit\'e Paris-Saclay, Universit\'e Paris Cit\'e, CEA, CNRS, AIM, 91191, Gif-sur-Yvette, France}

\author[0000-0002-0129-0316]{Savita Mathur}
\affil{Instituto de Astrof\'isica de Canarias (IAC), E-38205 La Laguna, Tenerife, Spain}
\affil{Universidad de La Laguna (ULL), Departamento de Astrof\'isica, E-38206 La Laguna, Tenerife, Spain}

\author{Benoit Mosser}
\affiliation{LESIA, Observatoire de Paris, Universit\'e PSL, CNRS, Sorbonne Universit\'e, Universit\'e de Paris, 92195 Meudon, France}

\author[0000-0002-4818-7885]{Jamie Tayar}
\affiliation{Department of Astronomy, University of Florida, Bryant Space Science Center, Stadium Road, Gainesville, FL 32611, USA }

\author[0000-0003-1479-3059]{Guy S. Stringfellow}
\affiliation{Center for Astrophysics and Space Astronomy, University of Colorado at Boulder, 389 UCB, Boulder, CO 80309-0389, USA}

\author[0000-0002-1691-8217]{Rachael L. Beaton}
    \affil{Space Telescope Science Institute, Baltimore, MD, 21218, USA}
    \affil{Department of Physics and Astronomy, Johns Hopkins University, Baltimore, MD 21218, USA}

\author[0000-0002-4912-8609]{Henrik J\"onsson}
\affil{Materials Science and Applied Mathematics, Malm\"o University, SE-205 06 Malm\"o, Sweden}

\author{Dante Minniti}
\affiliation{Instituto de Astrofísica, Facultad de Ciencias Exactas, Universidad Andres Bello, Av. Fernández Concha 700, Santiago, Chile e-mail: vvvdante@gmail.com}
\affiliation{Vatican Observatory, V00120 Vatican City State, Italy}


\begin{abstract}

We present a catalog of fundamental stellar properties for 7,673 evolved stars, including stellar radii and masses, determined from the combination of spectroscopic observations from the Apache Point Observatory Galactic Evolution Experiment (APOGEE), part of the Sloan Digital Sky Survey IV (SDSS), and asteroseismology from K2. The resulting APO-K2 catalog provides spectroscopically derived temperatures and metallicities, asteroseismic global parameters, evolutionary states, and asteroseismically-derived masses and radii. Additionally, we include kinematic information from \textit{Gaia}. We investigate the multi-dimensional space of abundance, stellar mass, and velocity with an eye toward applications in Galactic archaeology. The APO-K2 sample has a large population of low metallicity stars ($\sim$288 at [M/H] $\leq$ $-$1), and their asteroseismic masses are larger than astrophysical estimates. We argue that this may reflect offsets in the adopted fundamental temperature scale for metal-poor stars rather than metallicity-dependent issues with interpreting asteroseismic data. We characterize the kinematic properties of the population as a function of $\alpha$-enhancement and position in the disk and identify those stars in the sample that are candidate components of the \textit{Gaia-Enceladus} merger. Importantly, we characterize the selection function for the APO-K2 sample as a function of metallicity, radius, mass, $\nu_{\mathrm{max}}$, color, and magnitude referencing Galactic simulations and target selection criteria to enable robust statistical inferences with the catalog.

\end{abstract}


\section{Introduction} \label{sec:intro}

Galactic archaeology probes the stars in our Galaxy as a fossil record, investigating their histories to determine the formation and evolution of the Milky Way. Studying the Milky Way's fossil record is best achieved using precise stellar ages, abundances, and kinematics; this data reveals the \emph{when}, \emph{what}, and \emph{where} of the Milky Way's formation. 

Galactic archaeology gains the most from extensive data sets, with large numbers of stars over a broad parameter space, as ideally, these data sets represent the Galaxy's stellar population. Galactic stellar population research has experienced considerable growth in the last decade, with an abundance of space- and ground-based telescopes providing asteroseismic, spectroscopic, and kinematic measurements with which to determine ages, compositions, and positions for hundreds of thousands of stars; such data allows for a cogent description of the Milky Way's stellar building blocks for the first time. 

In terms of stellar targets, red giant stars are particularly beneficial for Galactic archaeology \citep[e.g.,][]{Stello2015}. Due to their high luminosities, red giant stars can be seen to greater distances than less evolved stars, providing a better understanding of the edge of our Galaxy; furthermore, their solar-like oscillations are observationally possible in a longer cadence than that which is necessary to observe similar oscillations in dwarfs and subgiants. Asteroseismic, spectroscopic, and kinematic catalogs of giant stars already exist for various samples. 

CoRoT \citep{Baglin2006}, \textit{Kepler} \citep{Borucki2010}, K2 \citep{Howell2014} and TESS \citep{Ricker2014} have been transformational for taking the once-boutique field of asteroseismology into the era of large data sets. There are catalogs of comprehensive asteroseismic mass data for red giants available from \textit{Kepler} \citep[$\sim$16,000][ see also \citet{Pinsonneault2018}]{Yu2018} and TESS ($\sim$158,000; \citet{Hon2021} and $\sim$1,700; \citet{Mackereth2021}). Other asteroseismic data sets have published stellar parameters but without spectroscopic cross-matching; for example, in TESS \citep[$\sim$1,700][]{Hon2022} and K2 \citep[$\sim$19,000, K2-GAP (DR3)][]{Zinn2022}, the latter of which we use in this work.

Spectroscopic data provide temperatures, gravities, and detailed chemical abundances. These data, for millions of stars, are being accumulated by APOGEE \citep{Majewski2017}, GALAH \citep{Buder2021}, and LAMOST \citep{Cui2012}. 

Finally, \textit{Gaia} \citep{Gaia2016} has also been ground-breaking in producing precise astrometric and kinematic information for millions of stars. 

We can learn \emph{when} stellar populations formed in our Galaxy using precise asteroseismic stellar ages \citep[e.g.,][]{Miglio2009,Sharma2016,SilvaAguirre2018a,Rendle2019,Mackereth2020b}. Such age estimates rely on asteroseismic masses in combination with stellar models, metallicities, and temperatures. When performing population asteroseismology on large samples of solar oscillators, it is common to depict the oscillation pattern with two global seismic parameters for faster processing instead of performing detailed mode-by-mode analysis on all targets. These asteroseismic global parameters are the large frequency separation ($\Delta\nu$; related to the mean density) and the frequency of maximum power ($\nu_{\mathrm{max}}$; related to $\log$(g) and T$_{\mathrm{eff}}$). These global parameters can, in turn, provide the input to calculate high-precision stellar masses and radii using scaling relations \citep[See Section \ref{sec:sample_data_astero} and][]{Kjeldsen1995, Brown1991}. These scaling relations, and resulting ages, depend explicitly on temperature measurements; therefore, spectroscopic surveys provide a powerful complement to time-domain asteroseismology.

Spectroscopic data measures stellar abundances powerfully describing \emph{what} the Galaxy's stellar composition is. For example, \citet{Zinn2022} explored the origin of neutron-capture elements for the K2 red giant sample, combining asteroseismology with abundances from GALAH, and used them to test their assumed production mechanisms by comparing them to models from \citet{Kobayashi2020}. An avenue where asteroseismic and spectroscopic measurements show symbiosis is the research of $\alpha$-capture abundances. The consensus states that as stars formed in our Galaxy, the proportion of metals increased through a series of supernovae, providing an increasingly metal-rich medium for star formation and altering the proportions of $\alpha$ elements to other elements (hereafter, [$\alpha$/Fe]) \citep[e.g.,][]{Burbidge1957,TimmesWoosleyWeaver1995}. Therefore, the most metal-poor old stars present an excellent probe of the nucleosynthesis pathways in the early Galaxy. Studies of the evolution of $\alpha$ element abundances can reveal much about the past epochs of star formation \citep{Kobayashi2020}. A major prediction from early models is a single-valued function of $\alpha$ elements as a function of [Fe/H] \citep{Haywood2016}. However, with the support of spectroscopic measurements, data have shown a double-valued function \citep{Fuhrmann1998,Prochaska2000,Bensby2003}. There are many proposed mechanisms for this [$\alpha$/Fe]-[Fe/H] bimodality, such as radial migration \citep{Sellwood2002,Schonrich2009,Nidever2014,Weinberg2017,Sharma2021}, two separate episodes of star formation \citep{Chiappini1997,Spitoni2019,Lian2020}, and stars forming throughout the Galaxy in clumpy bursts \citep{Clarke2019}. Ages from asteroseismology are unique tools allowing us to test the models of [$\alpha$/Fe]-[Fe/H] bimodality thanks to their precision, large sample sizes, and a relatively large range of distances compared to isochronal aging techniques. Samples from \textit{Kepler}, K2, and \textit{TESS} are promising for continued work to constrain these models \citep[e.g.,][]{SilvaAguirre2018a,Rendle2019,Mackereth2019b,Warfield2021}.

Because stars differ in their formation histories in the Galaxy, kinematic information is crucial to understand \emph{where} stars formed. With stellar position and kinematic energy values derived from astrometry, we can determine where stars formed and to which  Milky Way component they belong (the halo, thick disk, thin disk, or bulge); we can also determine if the stars originated as part of an accretion event by considering the Galactic halo. As the Milky Way developed its structure, it accreted much smaller galaxies, and in some cases, they are still actively accreting. For example, the tidal disruption of the Sagittarius dwarf galaxy is ongoing \citep{Ibata1994}, the Magellanic Clouds are on their first infall \citep{Besla2007}, and there are dozens of globular cluster streams currently encircling our Galaxy \citep{Bonaca2020}. These accreted bodies come with unique kinematic profiles, and they share integrals of motion \citep{HelmiDeZeeuw2000, Font2011, Simpson2019} even several billion years later, allowing us, in some cases, to identify their stellar components. These stars also share chemical abundance values, aiding in their identification \citep{Freeman2002, Venn2004, Lee2015, Grunblatt2021}. 

Combining the abundances, masses, radii, velocities, and positions of evolved stars allows us to assemble a catalog containing the essential data to address the significant questions in Galactic archaeology. Furthermore, combining these parameters can improve the precision of other diagnostic criteria. For example, evolutionary states can be inferred to higher precision by correlating spectroscopic properties with asteroseismic evolutionary \citep{Jonsson2020}, as it is otherwise difficult to separate shell H-burning (RGB) stars from core He-burning (RC) stars. It is also challenging to infer parameters for red giant branch (RGB) stars, such as ages, with spectroscopic information alone \citep[e.g.,][]{soderblom2010}.

Another problem our extensive data set can explore is the conflict between astrophysical priors and asteroseismic masses for low-metallicity stars in the halo \citep{Epstein2014}. Asteroseismic masses for main sequence and RGB stars are both accurate and precise to around $\sim$5\% \citep[e.g.,][]{Huber2017a,Zinn2019a,Li2020} in the near--solar metallicity regime. However, at low metallicities ([Fe/H] $=$ $-$1), asteroseismic masses of RGB stars are consistently larger than the values expected for old halo stars. Overestimates masses at low metallicities
has been explored in \textit{Kepler} data \citep{Pinsonneault2014}; however, the sample of such stars was small, and more data is needed to understand the origin of this effect. 

Our catalog introduces the first APOGEE and K2 combination and includes the incorporation of \textit{Gaia} data. Because the K2 fields sample very different populations from those seen in the original \textit{Kepler} field, we can gain powerful insights into the formation history of the Milky Way. This paper presents the results from the dedicated targeting efforts of SDSS-IV \citep{Beaton2021} to observe the K2 fields. The resulting APO-K2 catalog contains the combined data sets for 7,673 evolved Milky Way stars, combining spectroscopic \citep[APOGEE DR17,][]{Accetta2021}, asteroseismic \citep[K2-GAP,][]{Stello2015}, and astrometric \cite[\textit{Gaia} EDR3,][]{Gaia2021} data, with a well-understood selection function \citep[][hereafter S22]{Sharma2022}. In Section \ref{sec:sample_data}, we discuss the data and construction of the catalog. Section \ref{sec:final_sample} presents the final sample, including a discussion of metallicity, the selection function, and a comparison between the APO-K2 and APOKASC2 \citep{Pinsonneault2018} samples. In Section \ref{sec:discussion}, we investigate stellar masses in the low metallicity regime, the kinematic properties, and the [$\alpha$/Fe]-[Fe/H] bimodality as seen in the data set; we also identify halo stars and potential GES members \citep{Helmi2018, Mackereth2019}. Section \ref{sec:conclusion} presents the conclusions. The publication of asteroseismic ages for this sample will follow in a companion paper (Warfield et al. in prep.), as will a detailed analysis of sample abundances and multiplicity (Schonhut-Stasik et al. in prep.). 

\break 

\section{Sample Data}
\label{sec:sample_data}

Our catalog uses time domain data from the K2 Galactic Archaeology Program \citep[K2-GAP DR3,][]{Zinn2022}, spectroscopic data from APOGEE DR17 \citep{Accetta2021}, and astrometric data from \textit{Gaia} EDR3 (included in the APOGEE DR17 release \citep{Accetta2021}). Our initial sample is compiled by cross-matching K2-GAP DR3 with APOGEE DR17, keeping any source that appears in both (Section \ref{sec:sample_cross_match}). Having combined these sources we calculated fundamental asteroseismic parameters (Section \ref{sec:sample_data_astero}), incorporated spectroscopic parameters (Section \ref{sec:sample_data_spec}), and collated astrometric values (Section \ref{sec:sample_astrometric_data}), resulting in the APO-K2 catalog. 

\subsection{Cross-match}
\label{sec:sample_cross_match}

The K2-GAP DR3 catalog is the base of our cross-match, which has been analyzed in various works \citep{Stello2017, Zinn2020, Zinn2022}. The complete list of K2-GAP targets observed by K2 contains 121,715 stars. From here, a cross match with APOGEE DR17 was performed. Around 32\% of the K2-GAP stars were also observed in APOGEE (39,319 stars); by then cross-matching those stars with accurate asteroseismology, we have a sample of 8,581 stars (7\% of those observed for K2-GAP). 

The primary axis for our cross-match between the K2-GAP and APOGEE DR17 catalogs is their associated 2MASS IDs, which come from the Ecliptic Plane Input Catalog \citep[EPIC,][]{HUber2016}, compiled to support target selection and management for the K2 mission. Prior to the match, we sort the APOGEE DR17 targets by signal-to-noise ratio (SNR), dropping the rows with lower SNR where multiple entries for the same target are available. Dropping low SNR entries ensures that each entry in the K2-GAP catalog receives a single spectroscopic match, with the highest available SNR observation. To show which stars were dropped, we created a flag for the cross-match (\texttt{apo\_crossmatch\_flag}) and set the value to \texttt{1} where no drop was made and \texttt{0} otherwise. A minority of stars were observed in multiple K2 campaigns (10\% of the sample), and are noted in the final catalog as having observations in multiple campaigns. \textit{Gaia} EDR3 data is already in the APOGEE DR17 table, and so no stars are dropped by adding this information. 

\subsection{Asteroseismic Data}
\label{sec:sample_data_astero}

Asteroseismic detections for thousands of stars became possible with missions such as CoRoT \citep{Auvergne2009, DeRidder2009} and \textit{Kepler} \citep{Huber2009,Stello2013}. However, these missions had limited sky coverage, selecting stars primarily for planet-hunting \citep{Sharma2016, Sharma2017}. In \textit{Kepler}, this resulted in only a fraction of the available oscillating targets observed, at a given combination of age, metallicity, and distance \citep{SilvaAguirre2018a}.

The \textit{Kepler} mission targeted a field in the Cygnus and Lyra constellations, with a four-year nominal mission, to find Earth-like planets around Sun-like stars. Conveniently, \textit{Kepler's} precise photometry also allowed for the study of asteroseismic variability in thousands of cool stars; the long cadence of \textit{Kepler} being particularly advantageous, with 29.4-minute observations ideal for studying the oscillation periods of red giants \citep[e.g.,][]{Bedding2010}.

Following the failure of the second of \textit{Kepler's} four reaction wheels in May 2013, the \textit{Kepler} spacecraft began on a second mission, K2. The spacecraft was re-purposed using solar wind for partial stabilization, allowing the telescope to continue to observe, although the acquisition of \textit{Kepler's} original field was no longer possible due to the lack of stable pointing. Therefore, unlike the fixed original field, K2 observed 18 regions across a 360$^{\circ}$. ecliptic field of view for $\sim$80 days at a time \citep{Howell2014}. Each K2 campaign covers 115.64 sq. deg. --- over 21 CCD modules, each made of 1024 $\times$ 2200 pixel CCDs (2\farcs{98} pixel scale); there are slight gaps between the CCDs and between each of the 21 modules. 

The K2 mission had several scientific objectives proposed through guest observer programs. The international collaboration of K2-GAP \citep{Stello2015} created a dedicated program to target red giants far beyond the solar neighborhood, which provided a trove of new asteroseismic data by detecting oscillations, with an aim to study the formation and evolution of the Milky Way. Roughly 25\% of the observed K2 targets were allocated to K2-GAP. 

K2 performed 18 full observing campaigns before a lack of fuel forced retirement in 2018. In \citet[][hereafter Z22]{Zinn2022}, the authors present K2-GAP for campaigns (C) C1-C8 and C10-C18\footnote{Some differences between campaigns are worth noting: C3 had a slightly shifted field of view due to a late change in roll angle; therefore, some of the proposed targets were unobservable. During C10, a permanent failure of one of the CCD modules occurred, resulting in stars in this, and all subsequent campaigns, being observed partially, or not at all. C18 has very few seismic detections. C9 was not used as it was a dense field chosen for microlensing, nor C19, which had very few asteroseismic detections due to its short duration.}. The Z22 catalog represents the largest asteroseismic sample of red giants, with both $\nu_{max}$ and $\Delta\nu$, in the literature to date.

Our catalog includes K2-GAP asteroseismic data from C1-C8 and C10-C18, and was analysed by members of the \textit{Kepler} Asteroseismic Science Consortium (KASC)\footnote{\url{https://kasoc.phys.au.dk}}. The K2-GAP lightcurves were reduced and calibrated in a manner appropriate for asteroseismology (see Z22, where the interested reader can find a full explanation). Briefly, each star in the sample received analysis from six independent pipelines, which returned asteroseismic parameters. The final asteroseismic values are the average of the pipeline values, with an outlier rejection algorithm applied. For the stars observed in multiple campaigns, a weighted average of the averaged pipeline values from each campaign serves as the final set of parameters.

The final two asteroseismic parameters in the K2-GAP catalog are the frequency of maximum oscillation power ($\nu_{\mathrm{max}}$) and the large frequency separation ($\Delta\nu$); we only adopt stars that have a measurement of both these values in K2-GAP. We determine asteroseismic masses and radii by combining these parameters with scaling relations that use both spectroscopic temperatures (APOGEE DR17) and correction factors defined in Z22, referred to as the radius and mass coefficients ($\kappa_{\mathrm{M}}$ and $\kappa_{\mathrm{R}}$, respectively). Full details are in Z22, but we include the equations here for completeness:

\begin{equation}
    \begin{split}
    \frac{R}{R_{\odot}} \approx \bigg(\frac{\nu_{\mathrm{max}}}{\nu_{\mathrm{max},\odot}}\bigg)\bigg(\frac{\Delta\nu}{f_{\Delta\nu}\Delta\nu_{\odot}}\bigg)^{-2}\bigg(\frac{T_{\mathrm{eff}}}{T_{\mathrm{eff},\odot}}\bigg)^{1/2} \\
    \equiv \kappa_{R}\bigg(\frac{T_{\mathrm{eff}}}{T_{\mathrm{eff},\odot}}\bigg)^{1/2}
    \end{split}
\end{equation}

\begin{equation}
    \begin{split}
    \frac{M}{M_{\odot}} \approx \bigg(\frac{\nu_{\mathrm{max}}}{\nu_{\mathrm{max},\odot}}\bigg)^{3}\bigg(\frac{\Delta\nu}{f_{\Delta\nu}\Delta\nu_{\odot}}\bigg)^{-4}\bigg(\frac{T_{\mathrm{eff}}}{T_{\mathrm{eff},\odot}}\bigg)^{3/2} \\
    \equiv \kappa_{M}\bigg(\frac{T_{\mathrm{eff}}}{T_{\mathrm{eff},\odot}}\bigg)^{3/2}
    \end{split}
\end{equation}

In the above, $f_{\Delta\nu}$ represents a model-dependent correction to the observed $\Delta\nu$ values. Its value depends on, in part, the temperature and metallicity of the star. In K2-GAP DR3 (Z22), the EPIC served as the source for this temperature and metallicity. With APOGEE DR17 temperatures and metallicities in hand, we update $f_{\Delta\nu}$ using \texttt{Asfgrid} \citep{Sharma2016} with its low-mass low-metallicity extension \citep{Stello2022}, in accordance with the treatment described in Z22, including alpha-dependent corrections to the metallicities. In the APOKASC publications, the so far published K2-GAP work, and in this work, our asteroseismic parameters adopt the method found in APOKASC2, where we adopt an empirical calibration against fundamental data to set the $\nu_{\mathrm{max}}$ zero-point (see Z22 for more details). Our catalog includes masses and radii using $f_{\Delta\nu}$ and $\nu_{\mathrm{max}}$ values that have been updated since their initial calculation in Z22. For $\nu_{\mathrm{max}}$, the empirical evolutionary-state corrections from Z22 are removed and re-applied using the spectroscopic evolutionary states used here; the solar reference value, $\nu_{\mathrm{max,}\odot}$, for both red giant branch and red clump stars, is taken to be 3076$\mu\mathrm{Hz}$. $f_{\Delta\nu}$ values have similarly been recalculated with our new spectroscopic evolutionary states and updated APOGEE DR17 temperatures and metallicities. We include the updated $f_{\Delta\nu}$ and an associated flag, mass and radii coefficients, stellar masses and radii, and their associated errors in Table \ref{tab:table_description}. 

\subsection{Spectroscopic Data}
\label{sec:sample_data_spec}

Through detailed studies of the stars in the Milky Way, APOGEE \citep{Majewski2017} is unravelling the compositional form of our Galaxy. APOGEE uses a high-resolution (R $\sim$ 22,500), infrared spectrograph \citep{Wilson2019} operating in the H-band. For the Northern Hemisphere observations, APOGEE is mounted on the Sloan Foundation 2.5m telescope \citep{Gunn2006} at Apache Point Observatory, New Mexico. In the southern hemisphere the 40-inch Irénée DuPont telescope \citep{Bowen1973} at Las Campanas Observatory, Chile, houses the instrument.

APOGEE data are reduced, wavelength-calibrated, and co-added according to \cite{Nidever2015}. Spectroscopic parameters are calculated using the APOGEE Stellar Parameters and Chemical Abundances Pipeline \citep[ASPCAP;][]{Holtzman2015,GarciaPerez2016} and calibrated according to \cite{Holtzman2018}, with the model grids and the interpolation method described by \citet{Jonsson2020}. APOGEE primarily targeted K2 evolved stars (RGB and Red Clump (RC)) as they are intrinsically luminous with significant flux in the infrared; allowing high SNR observations at large distances. K2-GAP stars that were not already observed in the GALAH survey \citep{Buder2021} were prioritized, although overlapping stars at lower priority were also observed. About half of the targets observed by APOGEE reside in the disk of the Milky Way ($b$ $\leq$ 16$^{\circ}$), with the remaining targets split between the bulge and halo. Information on K2 object targetting can be found in \citet{Zasowski2017, Beaton2021} and \citet{Santana2021}; the latter two papers include information on the relative weighting of the different classes of targets.

In this work, we use APOGEE data obtained during the fourth phase of the Sloan Digital Sky Survey (hereafter SDSS-IV) \citep{Blanton2017} and analyzed in the seventeenth (and final) data release \citep{Accetta2021} of SDSS-IV (hereafter DR17). DR17 contains 675,000 APOGEE targets over an additional two years of SDSS observations, in both hemispheres, compared to DR16 \citep{Jonsson2020}. The cumulative nature of SDSS data sets means that DR17 contains a reprocessing of all data obtained, processed, and released in previous data releases. Although, many beneficial, and significant, updates appear in the DR17 release we only detail changes to the data release and reduction process relevant to this work. These changes primarily affect the abundances, because APOGEE uses the infrared flux method (IRFM) as an absolute standard for effective temperature, and asteroseismology for the absolute standard for $\log(g)$. For instance, ASCPAP was updated for DR17, with new sets of spectral synthesis grids including NTLE effects for Na, Mg, K, and Ca, which will be considered more in the upcoming APO-K2 abundance paper. Furthermore, \textit{Gaia} EDR3 data is incorporated into the DR17 data products, allowing us to include astrometry in the catalog. DR17 also includes eight value-added catalogs, including the ``SB2'' catalog of spectroscopic binaries \citep{Kounkel2021} and the ``RedClump'' catalog of RC stars (derived using a spectro-photometric selection).   

In the APOKASC samples \citep{Pinsonneault2014, Pinsonneault2018}, spectroscopic evolutionary states were employed to differentiate between the RC and RGB stars. For the purposes of this work, we adopt spectro-asteroseismic evolutionary states. These are calculated by a similar method as described in \cite{Warfield2021}, and they depend on recalculating the temperature-, surface gravity-, and abundance-dependent cut performed in \cite{Warfield2021}. Therefore, these spectroscopic evolutionary states rely on values of $T_{\rm eff}$, $\log{(\mathrm{g})}$, and elemental abundances from APOGEE DR17, and are trained on the asteroseismically-derived RGB and RC classifications from APOKASC3 (Pinsonneault et al., in prep). Ultimately, we assign stars as being on the RGB if their uncalibrated surface gravity\footnote{In the APOGEE catalog, the "SPEC" subscript marks the uncalibrated version of the parameter.} $\log{(g)}_{\rm SPEC} < 2.30$, or, where $\log{(g)}_{\rm SPEC} \geq 2.30$, if
\begin{equation}
    {\rm [C/N]} \times 10^{3} < 59.15 \ - \ 3.455 \ (155.1 {\rm [Fe/H]_{SPEC}} \ + \ \Delta T)
\end{equation}
In this equation, $\Delta T = T_{\rm eff}^{\rm (SPEC)} - T_{\rm ref}$ is the difference between a star's uncalibrated effective temperature and its `reference' temperature, calculated from a fit to the ridge-line of known RGB stars in the APOKASC3 data-set, which we define as
\begin{equation}
    T_{\rm ref} = 4427.18 \ - \ 399.5 {\rm [Fe/H]_{\rm SPEC}} \ + \ 553.17 (\log{(g)} \ - \ 2.5)
\end{equation}

Although these new evolutionary states show promise in that they are reliable in classifying the states accurately, there is a harder boundary in the $\log(g)$ space than we see when using spectroscopy-only derived evolutionary states. For example, there are few RC stars above a $\log(g)$ of 2.2 and this should be considered when dealing with stars in this domain.

break 
\subsection{Astrometric Data}
\label{sec:sample_astrometric_data}

The largest ever source of precise astrometric data comes from the \textit{Gaia} mission \citep{Gaia2016}, which was launched in 2013 and measured the six-dimensional spatial and velocity distribution of nearly two billion stars in the Milky Way. The astrometric information in this work comes from the early-release of the third data release of \textit{Gaia} (hereafter \textit{Gaia} EDR3 or EDR3 \citet{Gaia2021}), included in the APOGEE DR17 data release.

\section{Results}
\label{sec:final_sample}

In this section we begin by exploring the selection function (Section \ref{sec:comparison_selection_function}). We then detail the catalog provided in this work (Section \ref{sec:apo-k2-catalog}). In Section \ref{sec:sample_metallicity} we discuss the overall metallicity as a function of campaign, before finishing with an exploration of the sample in terms of evolutionary states (Sections \ref{sec:sample-overview} and \ref{sec:sub-populations}).

\subsection{Selection Function}
\label{sec:comparison_selection_function}

\begin{figure*}[h!]
\includegraphics[scale=0.45]{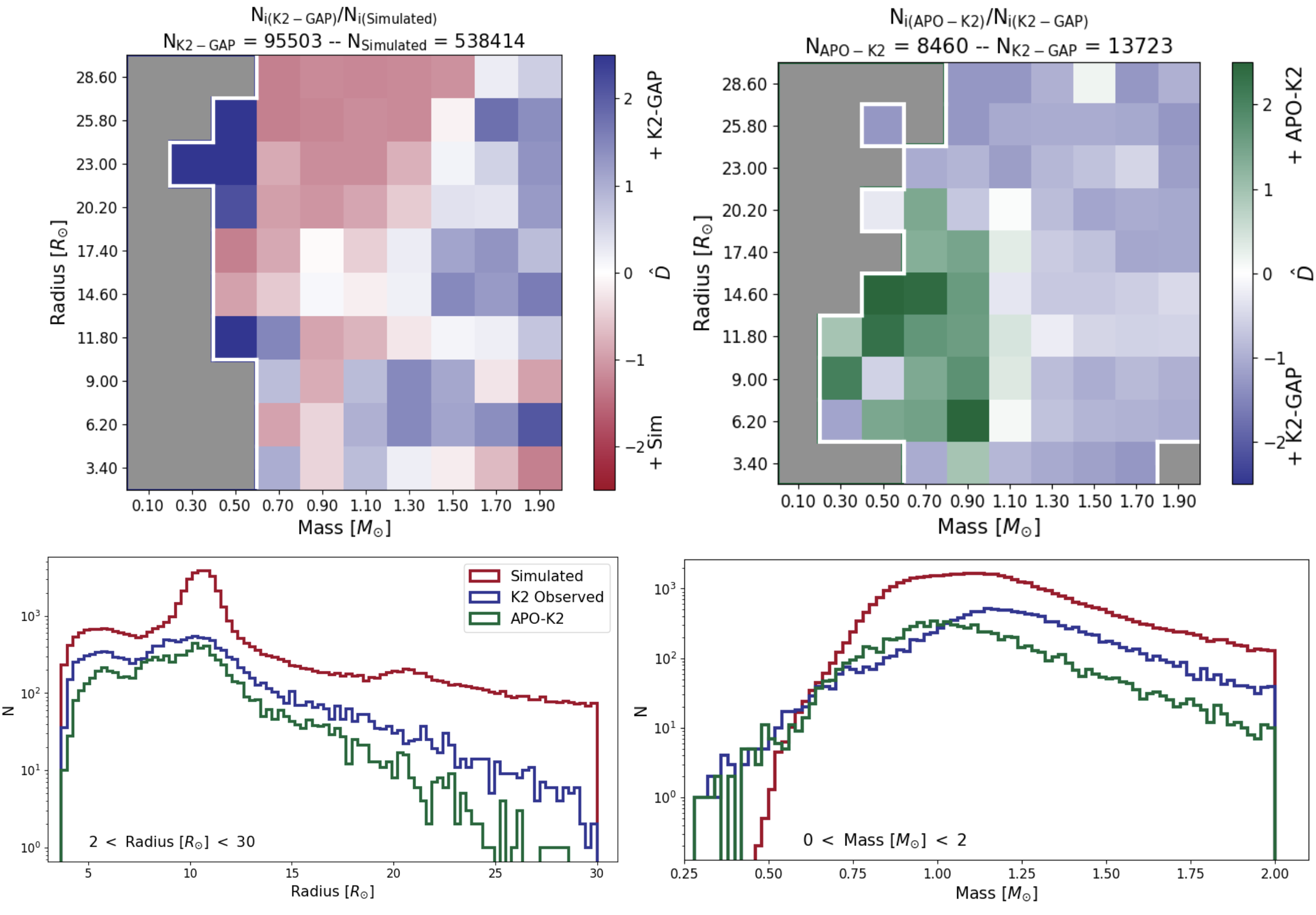}
\caption{A collection of plots describing the completeness of our sample through an analysis of the selection function, by first comparing the simulated and K2-GAP samples, then the K2-GAP and APO-K2 samples. \emph{Top Left:} A 10$\times$10 grid showing the fractional density of K2-GAP stars compared to the simulated sample in the mass [$M_{\odot}$] vs. radius [$R_{\odot}$] space. The title displays the number of stars in each of the samples overall; these values correspond to the scaling relation in the second term of Equation \ref{eq:scaling_d2}. The color-bar shows the varying density in each bin from an over abundance of simulated stars in red to an over abundance of K2-GAP stars in blue, compared to expectations. Because $\hat{D}$ contains an asinh scale, the color-bar is centered around 0, which corresponds to an equal number of stars in both bins. The grey area, bordered in white, delineates between bins with real numbers allocated and those that are calculated to have `nan' or `inf' values, due to the bin being populated by no star in either samples. \emph{Top Right:} The same as the top left but for the density of APO-K2 stars over K2-GAP stars, with an abundance of APO-K2 stars shown in green and an abundance of K2-GAP stars, again shown in blue. \emph{Bottom Left:} Three histograms showing the distribution in radius between 2 $<$ R [R$_{\odot}$] $<$ 30 of each sample (when cut on radius and mass); the simulated sample (red), the K2-GAP sample (blue), and the APO-K2 sample (green). The simulated stars have been multiplied by a scaling factor of 0.1, in all examples, to account for the over-sampling discussed in S22. \emph{Bottom Right:} The same as the bottom left but showing the distribution of masses over 0.25 $<$ M [M$_{\odot}$] $<$ 2. \label{fig:mass_rad_selection_plot}}
\end{figure*}

\begin{figure*}
\includegraphics[scale=0.40]{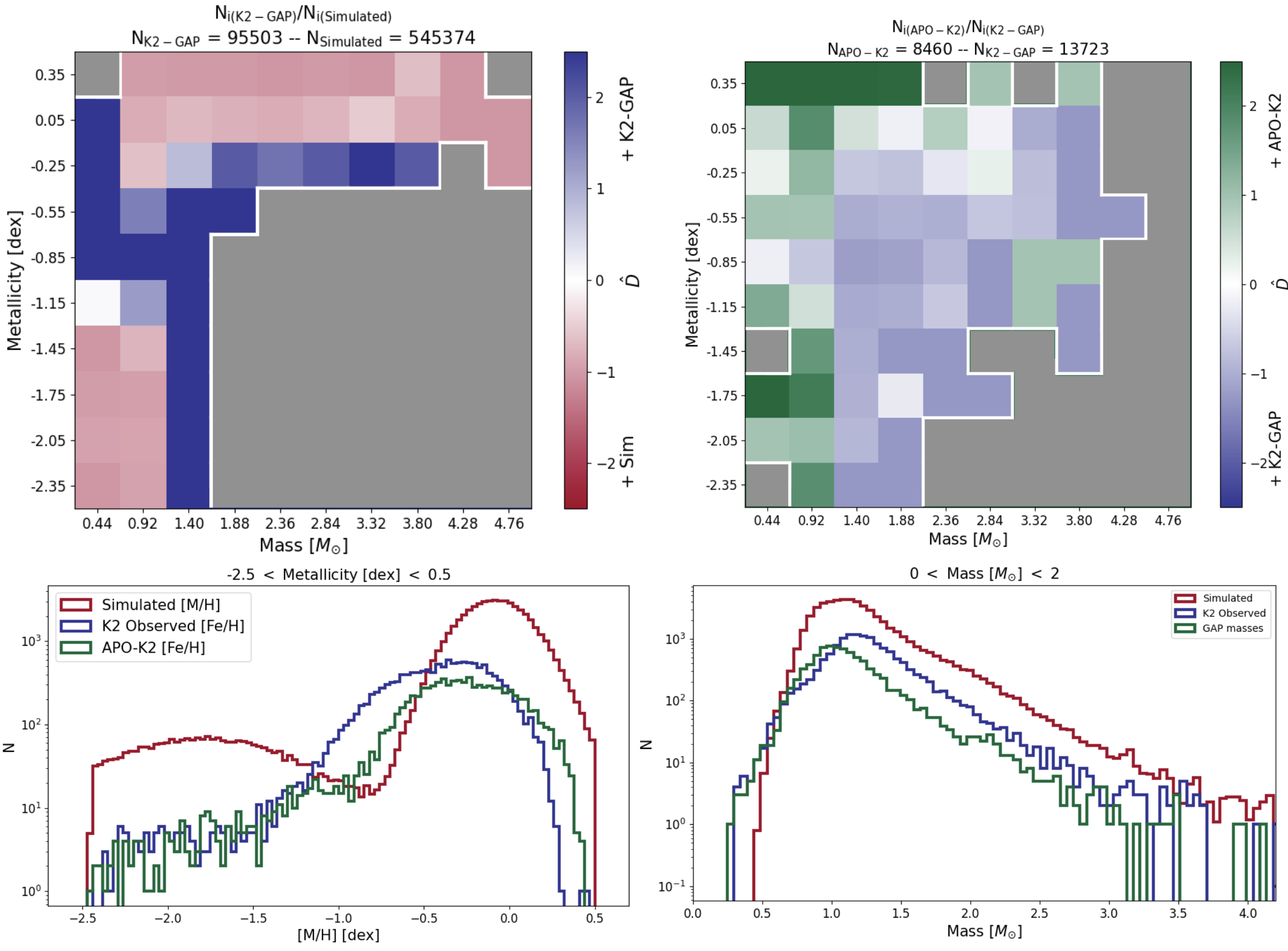}
\caption{Same as Figure \ref{fig:mass_rad_selection_plot} but in mass [$M_{\odot}$] vs metallicity [dex] space. The scaling values can be determined from the plot titles. The metallicity histogram (bottom left) ranges from -2.5 $<$ [M/H] [dex] $<$ 0.5, and the mass histogram (bottom right) ranges from 0 $<$ Mass [M$_{\odot}$] $<$ 2. See Figure \ref{fig:mass_rad_selection_plot} for more general information. \label{fig:mass_met_selection_plot}}
\end{figure*}

\begin{figure*}
\includegraphics[scale=0.40]{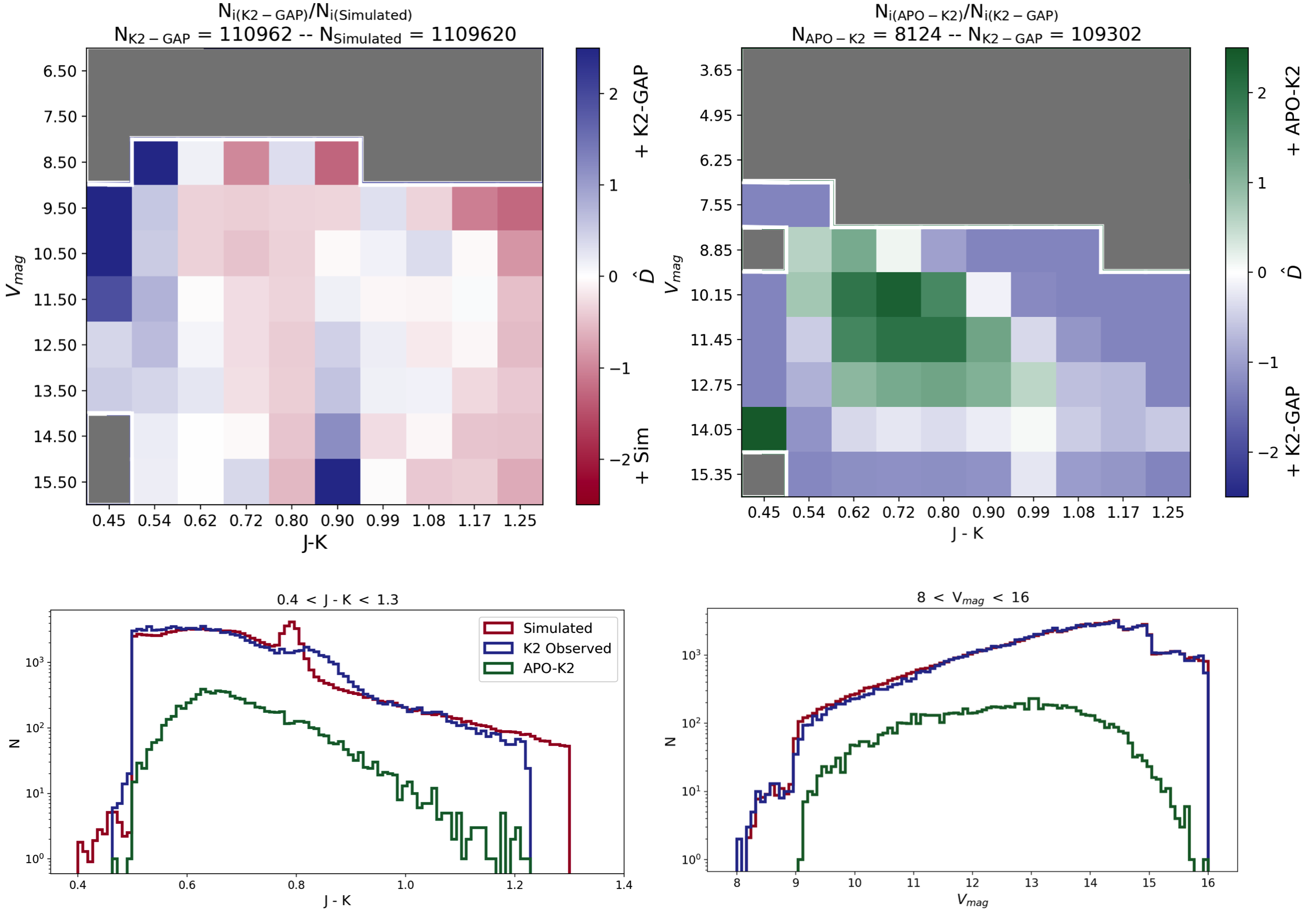}
\caption{Same as Figure \ref{fig:mass_rad_selection_plot} but in color (J - K) vs magnitude ($V_{mag}$). Again, the scaling value can be determined from the plot titles.  The color histogram (bottom left) ranges from 0.4 $<$ J - K $<$ 1.3, and the magnitude histogram (bottom right) ranges from 8 $<$ V$_{\mathrm{mag}}$ $<$ 16. See Figure \ref{fig:mass_rad_selection_plot} for more general information. \label{fig:color_mag_selection_plot}}
\end{figure*}

\begin{figure*}
\includegraphics[scale=0.50]{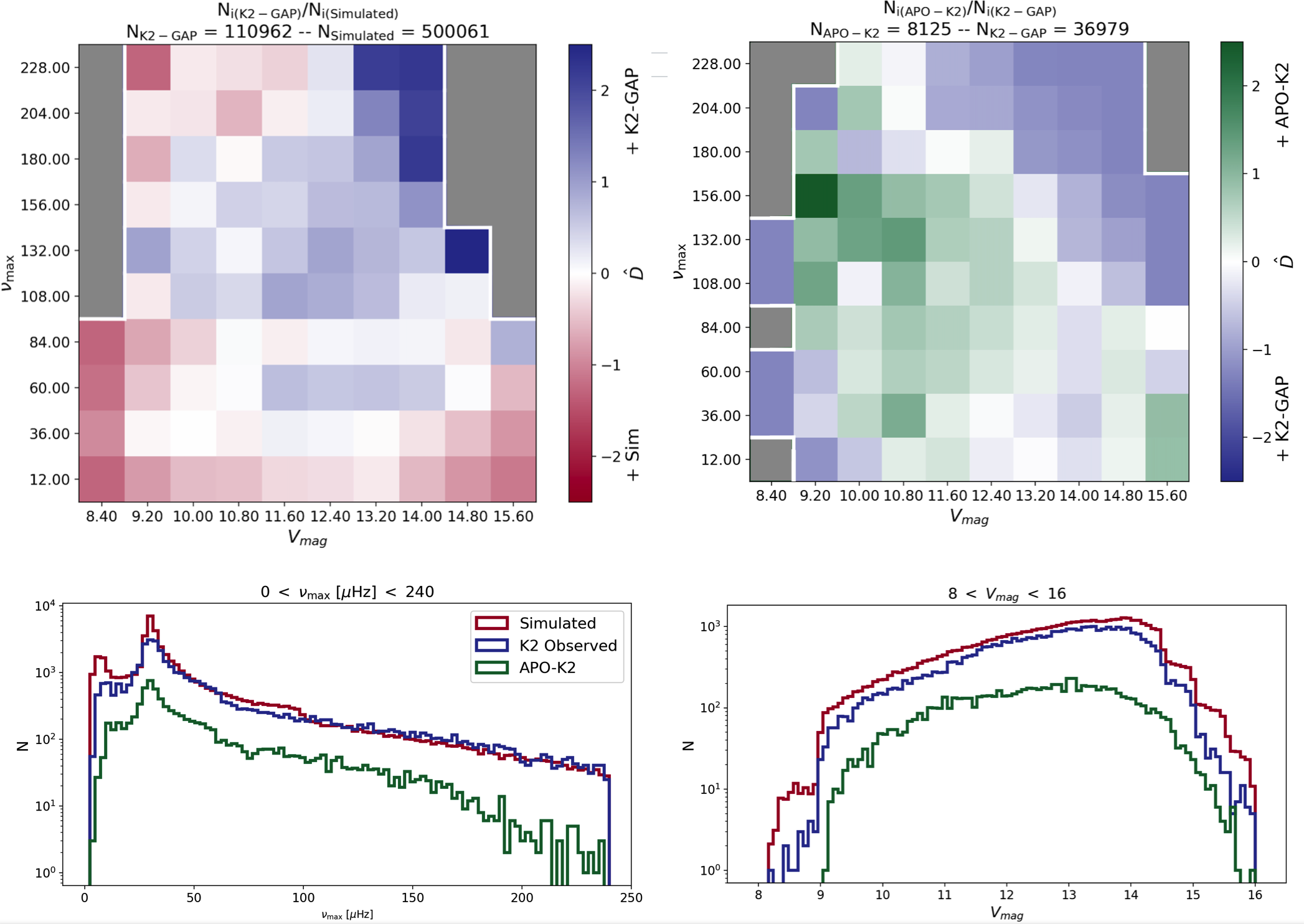}
\caption{Same as Figure \ref{fig:mass_rad_selection_plot} but in magnitude ($V_{mag}$) vs $\nu_{\mathrm{max}}$ [$\mu$Hz]. The scaling values can be determined from the plot titles.  The $\nu_{\mathrm{max}}$ histogram (bottom left) ranges from 0 $<$ $\nu_{\mathrm{max}}$ [$\mu$Hz] $<$ 240, and the magnitude histogram (bottom right) ranges from 8 $<$ V$_{\mathrm{mag}}$ $<$ 16. See Figure \ref{fig:mass_rad_selection_plot} for more general information. \label{fig:numax_mag_selection_plot}}
\end{figure*}

This section investigates the multiple layers of targeting and selection that lead to the APO-K2 sample, so that interested readers may consider completeness when using our catalog. The APO-K2 sample is fundamentally comprised of K2-GAP DR3, cross-matched with APOGEE DR17, which therefore includes selection and targeting choices made by both APOGEE DR17 and K2-GAP. In order to understand how our sample maps onto a distribution of the Galaxy's red giant population, we must first understand how these selection choices affect our data set. The K2-GAP data benefits from a simple and well-understood selection function (S22). The advantages of K2-GAP is the clear and reproducible target selection criteria, which can be used by us to map from K2-GAP to APO-K2.

Although the purpose here is to understand the completeness of the APO-K2 sample, we stop short of comparing it to any suggested Galactic distribution; instead we compare it to a theoretical target sample that was initially used in targeting for K2-GAP. Further comparison to underlying Galactic populations lie in the comparison between the \textit{Galaxia} models \citep{Sharma2011} used to simulated the sample and the accuracy of their underlying assumptions.

Some important assumptions are made in the creation of this model that could effect the way we perceive the completeness of our sample. For example, the assumed metallicity of the thin and thick disks (discussed further in our interpretation of Figure \ref{fig:mass_met_selection_plot}), and the choices related to star formation history (which directly impact the distribution of $\kappa_{\mathrm{M}}$, and as a consequence, mass). All choices made in the creation of the simulated sample and it's initial comparison to the K2-GAP targets are explored in \cite{Sharma2021}.

\subsubsection{K2-GAP Selection}

The targets for the K2-GAP sample were chosen with a color cut. Broadly, the cut removed mostly dwarfs, with (J - K$_\mathrm{s}$) $<$ 0.5, corresponding to dwarfs with $M_{\mathrm{K}\mathrm{s}}$ $>$ 1. This cut did exclude some giants, such as those on the horizontal branch (the blue extension of the RC); however, these are mostly rare, metal-poor stars, and are too hot to support solar-like oscillations. 

For C1, C2, and C3, the 2MASS H-band magnitude was used to select in brightness. For later campaigns, an approximation of the V-band magnitude measured from 2MASS J and K bands was used, as per Eq. \ref{eq:color_cut} below. This cut was chosen for later campaigns because K2 collects data in the $K_\mathrm{p}$ band, which is significantly bluer than the H-band, and more consistent with the V-band (see \citet{Sharma2022} for more details):

\begin{equation}
    V_{JK} = K_s + 2.0(J - K_s + 0.14) + 0.382 \exp{[(J - K_s - 0.2)/0.50]} \label{eq:color_cut}
\end{equation}

\subsubsection{APOGEE Selection}
By cross-matching with APOGEE we insert elements of the APOGEE selection function into our catalog. \citet{Beaton2021} discusses the prioritization scheme for the targeting in each field of the K2 program. This provides a meaningful comparison between targeted and observed stellar populations, and can be used to judge the completeness of our sample, in parameter spaces that may have been affected by our cross-match.

\subsubsection{Comparing Targeting and Selection Functions}
We compare our sample of APO-K2 stars to the observed K2-GAP stars and the simulated stars from S22\footnote{\url{github.com/sanjibs/k2gap} and \url{http://www.physics.usyd.edu.au/k2gap/} --- using \texttt{data\_name = `Galaxia-K2-sydai2-mrtd5'} for K2 observed stars and \texttt{data\_name = `k2-sydai2'} for the simulated stars.}. We note that the values of asteroseismic mass and radius that we report are the average over a number of pipelines from Z22, whereas the K2 data only use the SYD pipeline \citep{Huber2009}, which may cause a small difference in the inferred selection function when mapping from K2 data to APO-K2. 

From here we investigate four parameter spaces: mass vs. radius (Figure \ref{fig:mass_rad_selection_plot}), mass vs. metallicity (Figure \ref{fig:mass_met_selection_plot}), color vs. magnitude (Figure \ref{fig:color_mag_selection_plot}), and magnitude vs. $\nu_{\mathrm{max}}$ (\ref{fig:numax_mag_selection_plot}). In each space, depending on the relative density being computed, we cut down the larger sample to match the limits of the smaller sample\footnote{For example, when calculating the number density between the K2 and simulated data, we cut the simulated data to match the K2-GAP data limits.}. We created two 10 $\times$ 10 grids, in the investigated parameter space; for example, in the mass vs. radius regime we bin in a 10 $\times$ 10 grid of mass and radius. We populated these grids with the relative density of stars; one corresponding to the observed K2 sample/simulated sample, and one corresponding to the observed K2 sample/APO-K2 sample. This allows us to determine which stars are lost during the K2 observation phase and which from the APOGEE selection function. When considering the simulated sample we multiplied the amount of stars in each bin by 0.1, to compensate for how the simulated stars were over-sampled by a factor of 10 to reduce Poisson noise (see explanation in \citet{Sharma2022}). In all the density plots (Figures \ref{fig:mass_rad_selection_plot}, \ref{fig:mass_met_selection_plot}, \ref{fig:color_mag_selection_plot}, and \ref{fig:numax_mag_selection_plot}) grey areas cover bins where `nan' or `inf' values were calculated; this occurs when one of the samples has no stars in that bin. The resulting relative densities (D) were calculated and multiplied by a scaling factor (the second term in Equations \ref{eq:scaling_d1} and \ref{eq:scaling_d2}) that is specific to the parameter space. These scalings allow the reader to plainly see the differences in the bins. These equations are shown below, for the density of the K2 sample relative to the simulated sample: 
\begin{equation}
    \mathrm{D} = \frac{N_{i (\mathrm{K2})}}{N_{i (\mathrm{sim})}} \times \frac{N_{\mathrm{sim}}}{N_{\mathrm{K2}}} \label{eq:scaling_d1}
\end{equation}
and for the density of the APO-K2 sample relative to the K2 sample,

\begin{equation}
    \mathrm{D} = \frac{N_{i (\mathrm{APO-K2})}}{N_{i (\mathrm{K2})}} \times \frac{N_{\mathrm{K2}}}{N_{\mathrm{APO-K2}}} \label{eq:scaling_d2},
\end{equation}
where the first term represents the relative density between the samples for that bin. Another scaling was applied to all bins, to more easily see structure in the selection function (Equation \ref{eq:scaling_d_hat}). Using the value of $x$ = 0.6 (for the mass-radius space) and $x$ = 1.0 (for the other parameter spaces) in the denominator will tend the scaling toward a log-scale at $< -$0.6 and $>$ 0.6, and $<-$1.0 and $>$ 1.0, respectively. 

\begin{equation}
    \mathrm{\hat{D}} = \mathrm{arcsinh}\Bigg[\frac{D - 1}{x}\Bigg] \label{eq:scaling_d_hat}
\end{equation}

One of the key distributions, when considering Galactic archaeology, is mass. The mass distribution we would expect to see would be indicative of the most recent burst of star formation and the lifetime of the giant branch as a function of mass and metallicity \citep{Wu2017}; these factors conspire to produce the observed mass distribution. In terms of individual mass limits we expect a couple of clear boundaries. For example, no massive stars above 5 M$_{\odot}$. This is because high-mass red giants do not show solar-like oscillations and therefore are not observed in our sample, while their evolved counterparts have lower surface gravities than we probe. The lowest mass stars we expect would correspond to the age of the Galaxy, discussed further in Section \ref{sec:mass_vs_met}. Higher or lower masses than these bounds likely reflect mergers, binary mass transfer, or mass loss.

Figure \ref{fig:mass_rad_selection_plot} shows density plots in mass [$M_{\odot}$] vs. radius [$R_{\odot}$] and histograms for these parameters. The color-bar on the density plot indicates the scaled fractional density of stars. The scaling factor described in the second term of Equation \ref{eq:scaling_d1} for the left hand plot is (538414/95503) $\sim$ 5.6 (Number of K2-GAP stars/Number of simulated stars) and (13722/8460) $\sim$ 1.6 (Number of APO-K2 stars/Number of K2-GAP stars) on the right (Equation \ref{eq:scaling_d2}). In the top left density plot, for K2/simulated stars, we see few simulated stars in the low mass regime. This is likely due to the uncertainties on mass adopted by the simulation, as they correspond to the median uncertainties for the data \citep{Sharma2011}. We used the temperatures, $\kappa_{\mathrm{M}}$ values, a 3\% uncertainty on temperature (in alignment with EPIC temperature), and SYD $\kappa_{\mathrm{M}}$ errors to compute the fractional error on mass for the simulated stars. Over the whole sample, the fractional mass error for the simulated stars is around 24\%, but 51\% when the sample is cut to M $<$ 1M$_{\odot}$ stars. Therefore, as the simulation adopts median uncertainties from the data, the simulation errors for mass are underestimated. This likely causes the deficit in mass we see in the low-mass regime for the simulated sample; this is also seen at the low mass end of the mass histogram. 

Looking to the top right plot, of the relative APO-K2 stars/observed K2 stars, we see more APO-K2 stars at lower masses and radii (this is represented by the green bump on the left of the plot). This may be the result of underestimated masses, caused by different asteroseismic coefficients (and therefore temperatures) being used to calculate the masses and radii in these samples. To investigate the extent to which temperatures are the cause, we calculate the mean raw APOGEE temperatures and mean EPIC temperatures (used for the K2 sample). The resulting temperature difference corresponds to a difference of $\sim$6\% in mass. One could argue that the abundance of low-mass stars seen in the sample, that are absent in the simulated data, may be the result of mass loss (including mass transfer and binary interaction that the simulation does not account for), as discussed in works that investigate inferred mass loss from asteroseismic data \citep[][and Roberts et al., (in prep.)]{Miglio2012, Miglio2021, Tailo2022, Howell2022, Kallinger2018}, however due to the level of uncertainty in the simulated sample at low masses (as discussed above), we do not wish to draw this conclusion here. The same argument could be made for the K2-observed (green) histogram appearing to rise above the (APO-K2) blue histogram in the bottom right plot; although we would expect that in all cases the APO-K2 histogram should fall below the K2-observed histogram as cuts are being made to the K2-GAP sample when observed by APOGEE, the masses in these samples were calculated using different temperatures and will result in different masses, likely creating this effect.  We discuss further the importance of the temperature scale for asteroseismic masses in Section \ref{sec:mass_vs_met}. 

On the bottom left, in the radii histogram, we see a bump at 10R$_{\odot}$ that does not appear in the observed samples. This likely represents clump stars, which are underrepresented in the data, as it can be difficult to detect oscillations in RC \citep{Mosser2018} with their lower oscillation amplitudes \citep{Yu2018}. The extent of this bump in the histogram is due to the relatively long lifetime the stars spend in the clump relative to the RGB. RC stars share a similar core mass, which dictates their similar radii to one another (however they do not share the same total mass). 

Figure \ref{fig:mass_met_selection_plot} shows the density plots for mass [M$_{\odot}$] and metallicity [dex]. In this case, we used [M/H] values\footnote{When discussing abundances in this work we use the standard notation: $[\mathrm{X/Fe}] \equiv \log_{10}\bigg(\frac{\mathrm{X}}{\mathrm{Fe}}\bigg) - \log_{10}\bigg(\frac{\mathrm{X}_{\odot}}{\mathrm{Fe}_{\odot}}\bigg)$.} for the simulated stars, and [Fe/H] for APO-K2 and the observed K2 sample, as [Fe/H] was unavailable for the simulated sample\footnote{[M/H] is the average bulk metallicity in APOGEE, while [Fe/H] involves a selected subset of iron lines. In practice, the two agree closely for most APOGEE stars, since optical metallicity values track iron abundances (See \url{https://www.sdss3.org/dr10/irspec/aspcap.php} for more information.)}. For the relative density of the K2/simulated sample we see a bimodal distribution separated by an area where the relative amount of observed stars is higher. The bimodal distribution is likely due to the assumed metallicity of the thin and thick disks in the \texttt{Galaxia} model used to create the simulated stars. For example, the metallicity adopted in the model of the thick disk (-0.162) corresponds to the peak of the simulated histogram \citep{Sharma2019}. 

On the right hand side, showing the relative density of observed K2 stars/APO-K2, we see more colored bins, indicating the presence of more populated bins. This demonstrates that although these stars were observed in K2 they did not appear in the simulated sample, resulting in the grey area on the left hand density plot. 

In the mass histogram, again we see an wealth of lower mass stars ($\approx$ 0.6 $-$ 1 M$_{\odot}$) in the APO-K2 sample. In the right hand histogram we see that more stars were observed at lower masses in K2 than simulated. Note the difference in mass range in Figure \ref{fig:mass_rad_selection_plot} compared to Figure \ref{fig:mass_met_selection_plot}. The mass range in Figure \ref{fig:mass_rad_selection_plot} is purposefully more condensed to show the detail toward smaller masses. 

In Figure \ref{fig:color_mag_selection_plot}, we show the color-magnitude space. These plots clearly show color cuts enforced by the K2-GAP target selection, where C1 and C2 have a maximum magnitude of 7 in H-band, and all other campaigns have a maximum magnitude of $V_{\mathrm{JK}}$ $=$ 9 (see Table 1 of S22). In the top left density plot for the K2/simulated sample we see an abundance of K2 observed stars towards the bluer end of the color axis, and more simulated stars toward the red end. This is due to a number of factors: firstly, we would expect that as K2 reaches its magnitude limit it will be less able to see faint, red, stars that would be found in the simulation. Furthermore, the \texttt{Galaxia} simulation does not include extinction, which is a process left to the user, so this blue edge may be due to the cut off in J-K observation; therefore, the simulated sample is not seeing some stars that would be reddened and pass the J-K cut. This lack of extinction is also evident in the histograms, which show the simulated stars blueward of the K2 observed stars. Finally, the simulations assume a color dependence to the amplitude of oscillation, which may cause a mis-match between the assumed and actual color dependence of the amplitudes. 

Figure \ref{fig:numax_mag_selection_plot} shows the $\nu_{\mathrm{max}}$ vs. magnitude space. In both the density plots we see fewer K2 and APO-K2 stars at low magnitudes (brighter stars), likely due to targeting based on color-magnitude cuts (J -- K$_s$ $\geq$ 0.5 and 7 $\lesssim$ H $\lesssim$ 14) across the sky. We also see more K2 stars towards fainter stars with high $\nu_{\mathrm{max}}$ values. S22 plot their selection function in this same parameter space (bottom right hand plot of Figure 10 in S22), and the top left density plot in this figure shows similar results, confirming consistency in our work. In the density plot on the right, we see a shift in the distribution, imposed by the APOGEE selection. In the upper right corner of this plot, we see less APO-K2 stars, corresponding to less of the seismically low SNR stars.

The histograms between the K2 stars and the simulated stars match well in this set of plots because the selection of stars for the K2 sample were based on the ability to determine pulsations, and the simulated sample were selection-function-matched to the catalog (provided in S22) and designed to determine the completeness of the K2 observed stars. In the right hand histogram we see very good agreement between the simulated stars and the observed K2 stars. This is due to the selection function for targeting taking place in V-band, and these cuts being easily replicated in the simulation. The histograms also show an overabundance of stars at lower $\nu_{\mathrm{max}}$ values corresponding to the RC. 

The selection function plots for each of the campaigns can be found on the companion GitHub. It is important to consider the selection function over each campaign due to potential differences (discussed in S22), such as light curve duration, pointing accuracy, and the variation in crowding. For those interested in studying completeness in the sample, the supplemental material on the GitHub also includes tabulated values of the 2D histograms in XXX vs. YYY vs. ZZZ etc., where X is the bin edges for the x-axis, Y is the bin edges for the y-axis and Z is the relative density of stars in the bin.

\subsection{The APO-K2 Catalog}
\label{sec:apo-k2-catalog}
The public catalog distributed with this publication contains a row for each star, including the EPIC ID, K2 campaign number, and spectroscopic evolutionary state\footnote{Seismic evolutionary states for the K2-GAP sample can be found in] \cite{Zinn2022}}. From APOGEE, each row contains $T_{\mathrm{eff}}$ and $\log(g)$, with associated errors and [M/H], [Fe/H], [$\alpha$/M]. For K2, we include asteroseismic $\nu_{\mathrm{max}}$, $f_{\Delta\nu}$, and $\Delta\nu$, mass, radius, their asteroseismic errors, along with the mass and radius coefficients. We derived kinematic information from \textit{Gaia} positions, parallax, proper motions, and radial velocities. These kinematics include Galactic eccentricity, angular momentum, and total energy, derived using the \texttt{Python} module \texttt{Gala} \citep{Gala2017, GalaZenodo2020}. We also provide two flags, one for determining whether a source is high-[$\alpha$/M] or low-[$\alpha$/M], with an extra condition for stars that are close to the dividing line. The second flag pertains to the $f_{\Delta\nu}$, where 0.0 corresponds $f_{\Delta\nu}$ value within the bounds of the grid and with a complete [Fe/H], the value is 1.0 if 75\% of the Monte Carlo chains go outside the bounds of the $f_{\Delta\nu}$ grid, and 2.0 if there is incomplete [Fe/H], T$_{\mathrm{eff}}$, $\nu_{\mathrm{max}}$, or $\Delta\nu$ information to compute $f_{\Delta\nu}$.

Access to the APO-K2 catalog can be found as an electronic table with this paper and on the companion GitHub (\url{https://github.com/Jesstella/APO-K2}). The public catalog contains all information needed to re-create the plots in this paper. The GitHub and paper website (\url{https://www.jessicastasik.com/apo-k2}) also contain supplemental plots, including the selection function density plots for the individual K2 campaigns, and their density matrices as .csv files. Furthermore, in the interest of accessibility, alternative text\footnote{For those who may be blind or visually impaired.} for plots can also be found at these sources, as well as author information, and relevant conference presentations. The APOKASC2 data used in this paper can be found directly at \url{http://vizier.u-strasbg.fr/viz-bin/VizieR-3?-source=J/ApJS/215/19}, or by way of \cite{Pinsonneault2018}.

\subsection{Sample Metallicity}
\label{sec:sample_metallicity}

\begin{figure*}[h!]
\includegraphics[scale=0.75]{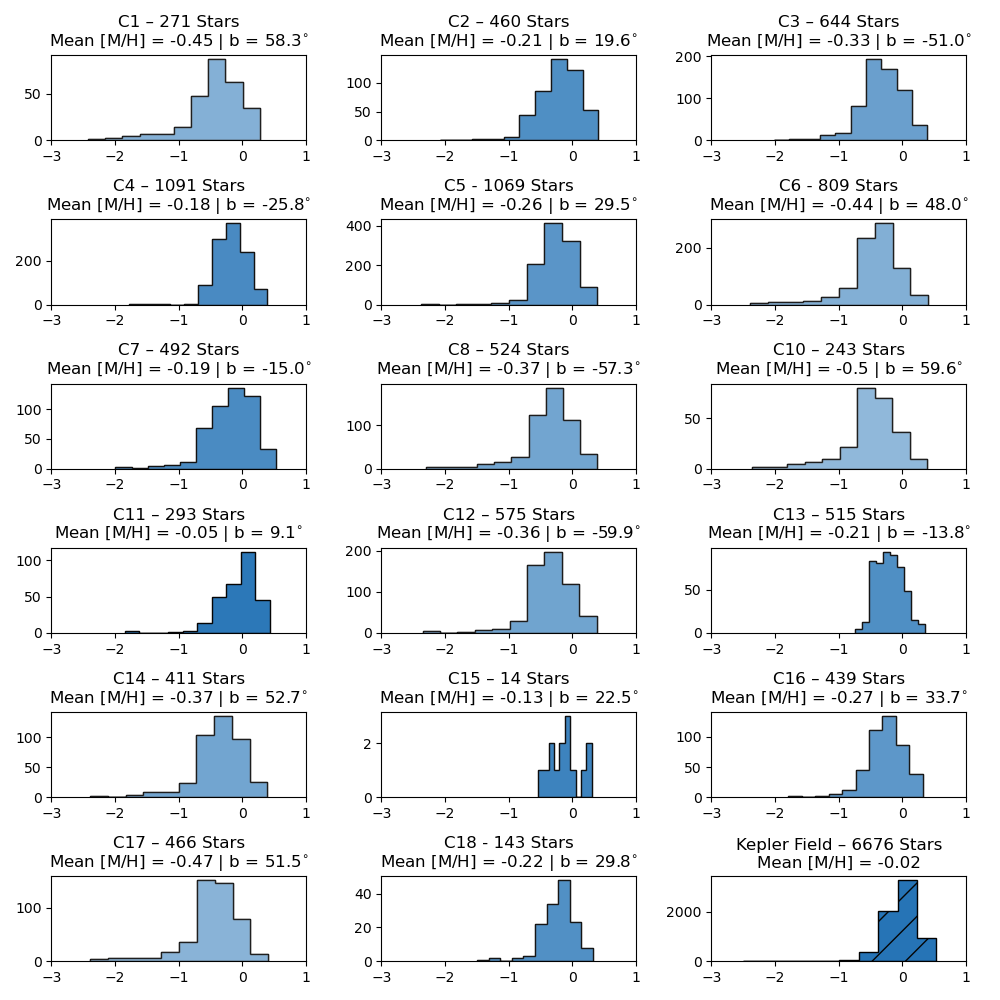}
\caption{Each histogram shows the APOGEE [M/H] distribution for the K2 campaigns. The bottom right-hand plot, with hatched markings, is the metallicity distribution of the \textit{Kepler} field taken from APOKASC2 \citep{Pinsonneault2018}. Color indicates the mean metallicity, with a darker blue corresponding to a higher mean metallicity. The title of each plot gives the campaign number, the number of stars in the campaign, the average metallicity, and the average Galactic latitude of the campaign field in. \label{fig:met_by_campaign}}
\end{figure*}

K2 samples multiple Galactic lines of sight providing a broad overview of the metallicity distribution of the Milky Way. Overall, our catalog provides a range of metallicities for evolved stars, particularly in campaigns of high and low Galactic latitudes. 

Figure \ref{fig:met_by_campaign} shows a histogram of [M/H] distribution for each K2 campaign. Each histogram contains all stars observed in the campaign, including those that appear in multiple campaigns. The bottom right histogram shows the metallicity distribution for the APOKASC2 sample. C10 boasts the lowest mean metallicity with $\langle$[M/H]$\rangle = -$0.50 [dex] at \textit{b} = 59.6$^{\circ}$\footnote{Where \textit{b} is the average Galactic latitude for the field.}. The highest average metallicity is in C11 ($\langle$[M/H]$\rangle = -$0.05 [dex] at \textit{b} = 9.0$^{\circ}$). This is in contrast to the APOKASC2 histogram with a mean metallicity of $-$0.02 [dex] (at \textit{b} = 13.5$^{\circ}$); all of the K2 histograms have lower mean metallicity than the \textit{Kepler} field.  

\textit{Kepler's} objective of observing near-by dwarf stars probably resulted in a dearth of metal-poor giants; therefore, displaying a striking difference to our sample. One factor in the difference in average metallicity is that many \textit{Kepler} giants were selected as planet candidate targets in a (generally) magnitude limited sample, strongly disfavouring luminous giants. This was then, in turn, biased against the metal-poor targets because they are (on average) further away, thus more likely to have a truly high luminosity at fixed magnitude.  Reinforcing the dearth of distant red giants, \citet{Wolniewicz2021} found a strong selection bias against cool, low-luminosity, red giant stars in \textit{Kepler}, where the observed red giants decrease from $\approx$ 80\% at $K_{\mathrm{p}}$ = 14 mag to $\approx$ 50\% at $K_{\mathrm{p}}$ = 15 mag; with only $40\%$ of red giants at $K_\mathrm{p}$ = 15 being observed for more than 8 quarters. They also note that the scarcity of observed red giants could be because the goal of \textit{Kepler} was to observe solar-type stars; therefore, many identified red giants were removed from the target list after one quarter. The K2-GAP sample was selected with completeness of the evolved stars in mind, a sample ideal for Galactic archaeology. Furthermore, the K2 sample is, on average, farther away from the Galactic plane than the stars in the \textit{Kepler} field. This is evident in Figures \ref{fig:sample_galaxy} and \ref{fig:met_by_campaign}, as the stars at higher and lower Galactic latitude are generally more metal-poor.

\subsection{Sample Overview}
\label{sec:sample-overview}

\begin{figure*}
    \centering
    \includegraphics[scale=0.68]{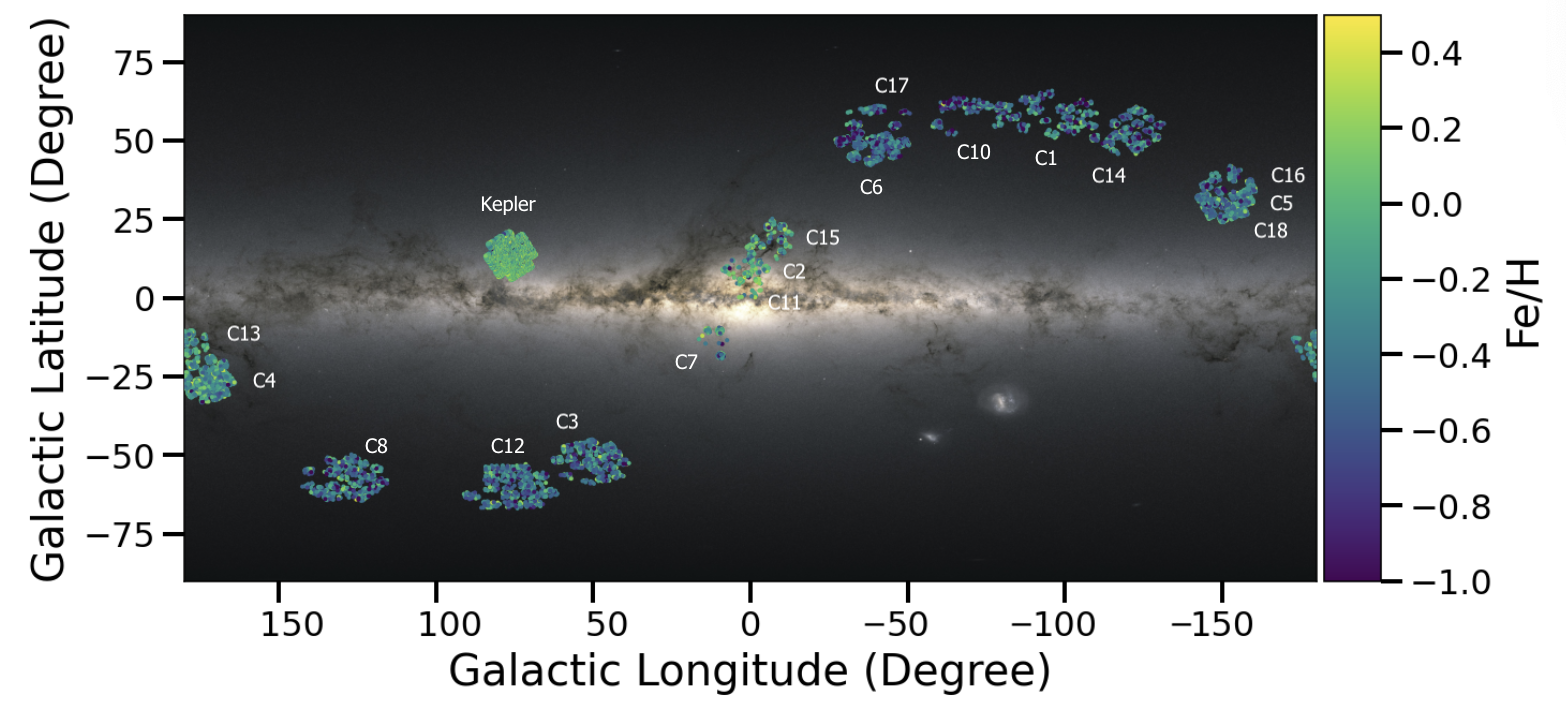}
    \caption{Footprint of each of the K2 campaigns in this sample, on a backdrop of the Milky Way, with axes representing the coordinates [degrees]. Each point represents a star in our catalog and color indicates its metallicity. The position of the \textit{Kepler} mission field is shown for reference, also colored by metallicity [Fe/H], as taken from APOKASC2. Gaps seen in the telescopes field of view correspond to CCD modules 3 and 7, which failed prior to the K2 mission. Each campaign is labelled with the campaign number in the format `C[number]'. The metallicity color-bar has been scaled between -1.0 and 0.5 to show the resulting metallicity distribution within each campaign. C9 and C19 are not used and therefore not shown. Background image modified from ESA/Gaia/DPAC, and is applied using the \texttt{mw\_plot} Python module and the \texttt{MWSkyProjection} map `equirectangular.' An accessible black and white plot with alternative text can be found at the companion GitHub and paper website.}
    \label{fig:sample_galaxy}
\end{figure*}

\begin{figure*}[h!]
    \includegraphics[scale=0.65]{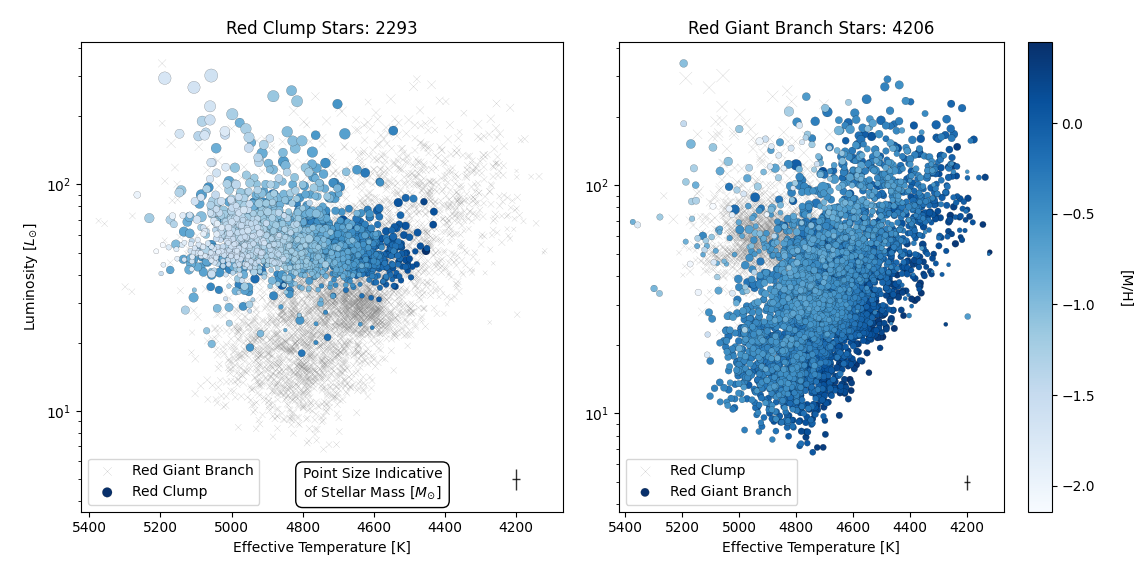}
    \caption{Two H-R diagrams showing the APO-K2 sample. \textit{Left:} RC stars. \textit{Right:} RGB stars. The title of each plot includes the number of stars in the plot. Only one observation of each star is plotted for stars that were observed in multiple campaigns. Stars with R $>$ 30$R_{\odot}$ are removed. The `ASPCAPFLAG'$^*$ must be 0 for the stars to be included in this plot. This removes 1,174 stars from the total sample. The marker size, for filled circles, corresponds to the asteroseismic stellar mass [$M_{\odot}$] (See Section \ref{sec:sample_data_astero}) and the color scale corresponds to [M/H]] [dex] from APOGEE. A representative error bar is given in the lower right of each plot for the stars shown as filled circles. The points are ordered by metallicity in descending order so that low metallicity stars are plotted on the top of the scatter. Luminosity is calculated using the asteroseismic radius and APOGEE temperature. Accessible versions of this plot can be found at the GitHub and website. \label{fig:hr_diagram}}
    \footnotesize\textsuperscript{*}{`ASPCAPFLAG' corresponds to a bit mask of warnings produced by the ASPCAP pipeline. The APOGEE bit masks are used to provide information associated with individual pixels in a one-dimensional spectrum (For more information, see \url{https://www.sdss4.org/dr17/irspec/apogee-bitmasks/}.) When a star has a value of a 0 integer from the bit mask, no warnings have been flagged for the observation.}
\end{figure*}

\begin{figure*}[h!]
    \includegraphics[scale=0.60]{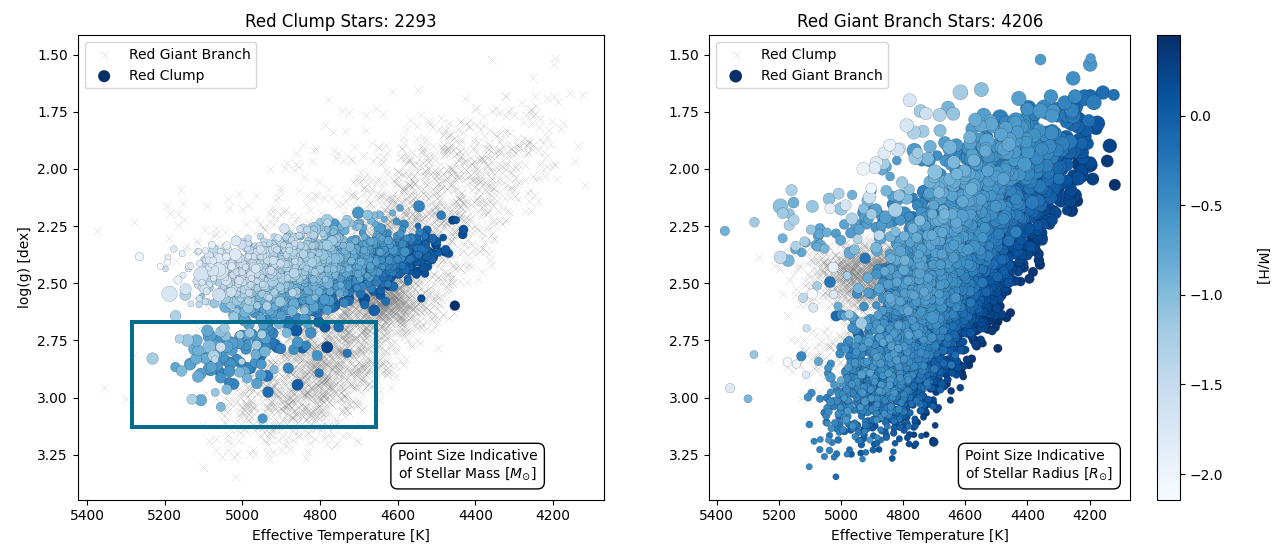}
    \caption{Two Kiel diagrams showing the APO-K2 sample. All cuts, color, and placement are the same as in Figure \ref{fig:hr_diagram} with the exception of the circle size for the RGB plot (right), which is indicative of asteroseismic stellar radius, as opposed to stellar mass. \label{fig:kiel_diagrams}}
\end{figure*}

\begin{figure*}
    \centering
    \includegraphics[scale=0.65]{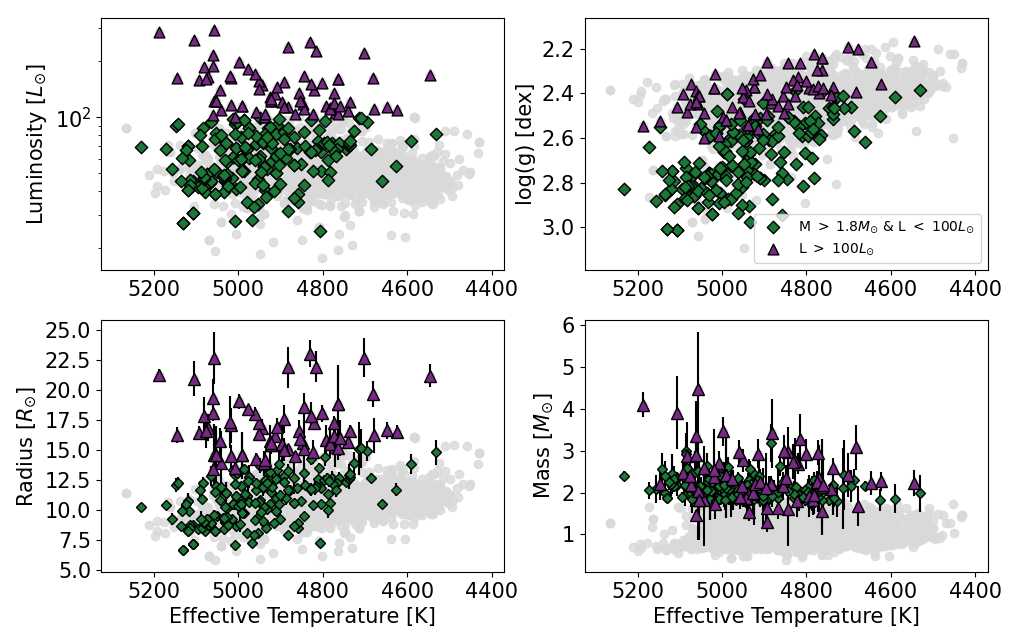}
    \caption{Four scatter plots showing the RC stars, in various parameter spaces. Each plot has T$_{\mathrm{eff}}$ on the x-axis and the y-axis depicting stellar Luminosity [L$_{\odot}$] (top left), $\log(g)$ [dex] (top right), Radius [R$_{\odot}$] (bottom left), and Mass [M$_{\odot}$] (bottom right). The entire sample of RC stars is shown by grey circles. Stars with L $>$ 100L$_{\odot}$ are represented by purple triangles and stars with L $<$ 100L$_{\odot}$ and M $>$ 1.8M$_{\odot}$ are shown by green diamonds. Error bars are shown for the radii and masses in the bottom left and right panels, respectively.}
    \label{fig:clump_plot}
\end{figure*}

Our final APO-K2 sample contains 7,673 unique stars with spectroscopic, asteroseismic, and astrometric data. The sample includes a total of 8,460 observations, occurring across multiple campaigns, sometimes observing the same star in multiple campaigns. When separated into evolutionary states, we have 2,465 unique RC stars and 5,208 unique RGB stars. The extensive overlap (See Section \ref{sec:sample_cross_match}) between the APOGEE catalog and K2-GAP program is due to the priority ordering for target selection (See Table 3 of S22). Targets were chosen for observation in K2-GAP according to criteria such as 2MASS and SDSS color and membership in existing spectroscopic survey catalogues, with each criterion given a priority ranking. This selection method gave the highest priority to APOGEE targets; APOGEE, in turn, prioritized observing spectra of K2-GAP stars resulting in the large overlap we see.

The sample reaches $\sim$60$^{\circ}$ above and below the plane of the Milky Way, and explores the Galactic Center. The position of these stars relative to the plane of the Milky Way are shown in Figure \ref{fig:sample_galaxy}. Color represents [Fe/H] for each star in our sample (from APOGEE), with the color-bar scaled between -1.0 and 0.5 [dex]; the actual maximum and minimum values of [Fe/H] are -2.45 and 0.51 [dex], respectively. This scaling shows the relatively metal-rich state of the \textit{Kepler} field in comparison to K2 and the metal-rich state of C2, C7, C11, and C15 relative to the other Campaigns; we discussed campaign-specific metallicities in Section \ref{sec:sample_metallicity}. We do not include C9, nor C19 (see Section \ref{sec:sample_data_astero}). Throughout this work we investigate, in particular, the low-metallicity stars specifically those below $-$1.0 [dex], which have not been available in large asteroseismic catalogs until now. For comparison, there are $\sim$ 288 stars in this catalog with [Fe/H] $\leq$ $-$1.0, and only $\sim$ 35 in the APOKASC2 catalog. 

\subsection{Identifying Stellar Sub-Populations}
\label{sec:sub-populations}
We separated the APO-K2 sample into two sub-samples (RC and RGB) using their evolutionary states (as described in Section \ref{sec:sample_data_spec}). Figure \ref{fig:hr_diagram} illustrates that combining asteroseismology and spectroscopy enables us to decipher areas on the H-R diagram containing interesting details like the RGB bump and secondary red clump stars \citep{Tayar2019}. Figure \ref{fig:hr_diagram} represents the stars on a H-R diagram. The left plot presents the RC stars prominently, with the RGB stars indicated by grey crosses in the background, and the right-hand side presents the RGB stars prominently, with the RC stars indicated by the grey crosses. By separating these two groups of stars, we hope to make the different samples clear to the reader. Figure \ref{fig:kiel_diagrams} displays the same stars as in Figure \ref{fig:hr_diagram} but in $\log$(g) $-$ T$_{\mathrm{eff}}$ space, also known as a Kiel diagram. The size of the points indicates the asteroseismically derived mass for Figure \ref{fig:hr_diagram} and the left hand plot of Figure \ref{fig:kiel_diagrams}. The right hand side of Figure \ref{fig:kiel_diagrams} has asteroseismically derived radius represented by point size. 

The size of our sample is clear in Figures \ref{fig:hr_diagram} and \ref{fig:kiel_diagrams}, and coneys a large collection of RC and RGB stars. These figures show a rich array of features that are qualitatively in agreement with the physics of RGB and RC phases of stellar evolution \citep{Cox1968,Kippenhahn2013,Girardi2016}. We mention some of these patterns here, and ongoing investigation explores to what extent the RC observations are consistent with models as a function of mass and metallicity. These plots indicate the importance of precisely derived evolutionary states (see Section \ref{sec:sample_data_spec}); there is overlap between higher mass RGB stars and RC stars, which are more precisely distinguished asteroseismically that by using classical H-R distinction. 

The combination of spectroscopic temperatures with well constrained asteroseismic radii (used to calculate luminosity) allows us to investigate small sub-samples of secondary RC and RGB bump stars. The RGB bump stars are clearly shown by an over density of grey crosses on the left-hand side of Figure \ref{fig:hr_diagram}, at a temperature of $\sim$4650K and 30$L_{\odot}$.   

When considering these plots as one sample, the RC stars are generally at higher temperatures than their RGB counterparts. The RC stars show an increase in mass with temperature and luminosity. Using mass as a proxy for age, this gradient implies youth at higher temperature and luminosity. A clear metallicity gradient can be seen with temperature, with lower metallicities corresponding to hotter stars. The existence of the secondary red clump is seen on the left-hand side of Figure \ref{fig:kiel_diagrams}, with a collection of stars around $\log(g)$ $\sim$2.75 dex and $T_{\mathrm{eff}}$ $\sim$5000 K. We explore possible secondary red clump stars further in Figure \ref{fig:clump_plot}, as they are not seen clearly in the left hand plot of Figure \ref{fig:hr_diagram}.

The RGB stars also show a gradient in metallicity towards higher temperature and luminosity, most clearly in Figure \ref{fig:kiel_diagrams}, where metallicity decreases toward the left of the branch. This gradient also corresponds to increased radii (see marker size in Figure \ref{fig:kiel_diagrams}). These relationships and the slight dispersion of points at higher luminosities (representing a wider range of temperatures for the more luminous stars) may be representative of AGB stars, although it is difficult to fully determine stars belonging to the AGB. Using MIST models we were able to rule out AGB stars at $\log(g)$ $>$ 2.0 with reasonable certainty, for low metallicity stars ([Fe/H] $<$ -1.5), and with relative confidence at solar metallicities. However, we found a $\log(g)$ cut alone does not rule out all AGB stars. 

Using this sample we attempt to make a distinction between secondary RC stars in different stages of evolution. In Figure \ref{fig:clump_plot} we further investigate possible members of the secondary red clump. This plot shows RC stars plotted using four different parameters as a function of temperature. Each of these plots uses grey circles to show the RC sample as a whole, purple triangles to indicate stars with a luminosity $>$ 100L$_{\odot}$, and green diamonds to indicate stars with a mass $>$ 1.8M$_{\odot}$ and a luminosity $<$ 100L$_{\odot}$. The 1.8M$_{\odot}$ cut was used to select for secondary clump stars, which do not undergo a helium flash and therefore show a range of luminosities due to their range of core masses \citep{Girardi2016}. In T$_\mathrm{eff}$ vs. luminosity we see a number of high luminosity stars above the RC, whilst the more massive stars are hidden within the RC. However, when plotted using $\log$(g), the high luminosity stars are inside the RC, and the high mass stars fall below it, indicating that these two samples are separate and distinct.

The high luminosity clump stars are likely secondary red clump stars due to their high mass, but may be later in their evolution compared to the other high-mass clump stars. We would expect their radii to expand with evolution as the core of the star contracts and heats up, consistent with the majority of them having radii $>$ 12.5 R$_{\odot}$.


On the other hand, the stars that may be earlier in their evolution (green diamonds) appear on the Kiel diagram as secondary red clump stars, falling below the main RC. These stars also have higher masses than the main clump, as seen in the bottom-right figure, but their radii are more consistent with the RC than the purple triangles, as they have not yet started to expand.

\subsection{Comparison to APOKASC2}

\citet{Pinsonneault2014} presented the initial combination of asteroseismic (\textit{Kepler}) and spectroscopic (APOGEE) data for 1,916 evolved stars in the first APOKASC catalog (hereafter, APOKASC1). They used the asteroseismic data to calibrate the relationships between parameters such as mass and age with spectroscopic observables. The second APOKASC release \citep[][hereafter APOKASC2]{Pinsonneault2018} looked at an additional 4,760 evolved stars (6,676 in total) with an empirical approach, combining asteroseismic measurements across different methodologies to calculate averaged values and reduce systematic errors. APOKASC1 used SDSS DR10 \citep{Meszaros2013} parameters, and the APOKASC2 used SDSS DR14 \citep{Holtzman2015}. 

The K2-GAP asteroseismic parameters used here follow a similar averaging approach \citep{Zinn2022} to APOKASC2. The main difference between the approachs is in target selection. The K2 stars were chosen =as a function of magnitude and color with the intention of creating a clean and easy to reproduce sample, unlike the \textit{Kepler} stars.

K2 is better suited to Galactic archaeology as compared to \textit{Kepler}. Due to its multiple Galactic lines of sight, it allows wider coverage of the Galaxy, observing multiple stellar populations at greater distances both radially and above and below the Galactic plane, thus broadening our understanding of the Galaxy's stellar composition as a whole. However, the downside to K2's wide coverage is the shorter length of lightcurves in comparison to \textit{Kepler}, resulting in lower SNR, meaning that oscillation spectra are harder to analyze.


APO-K2 and APOKASC2 also differ to APOKASC1 in the addition of stars in the low-metallicity regime (See Figure \ref{fig:met_by_campaign}). Furthermore, in APO-K2 we also observe stars across a wide range of Galactic latitude and longitude and to farther distances. These broader parameter spaces subsequently extend our understanding of related parameters, i.e. the high-mass vs. low-metallicity space (See Section \ref{sec:mass_vs_met}) and the [$\alpha$/Fe]-bimodality (See Section \ref{sec:alpha_bimodality}).

\section{Discussion: Example Applications of the APO-K2 Catalog}
\label{sec:discussion}

\subsection{Asteroseismic Mass vs. Metallicity}
\label{sec:mass_vs_met}

\begin{figure*}
\includegraphics[scale=0.45]{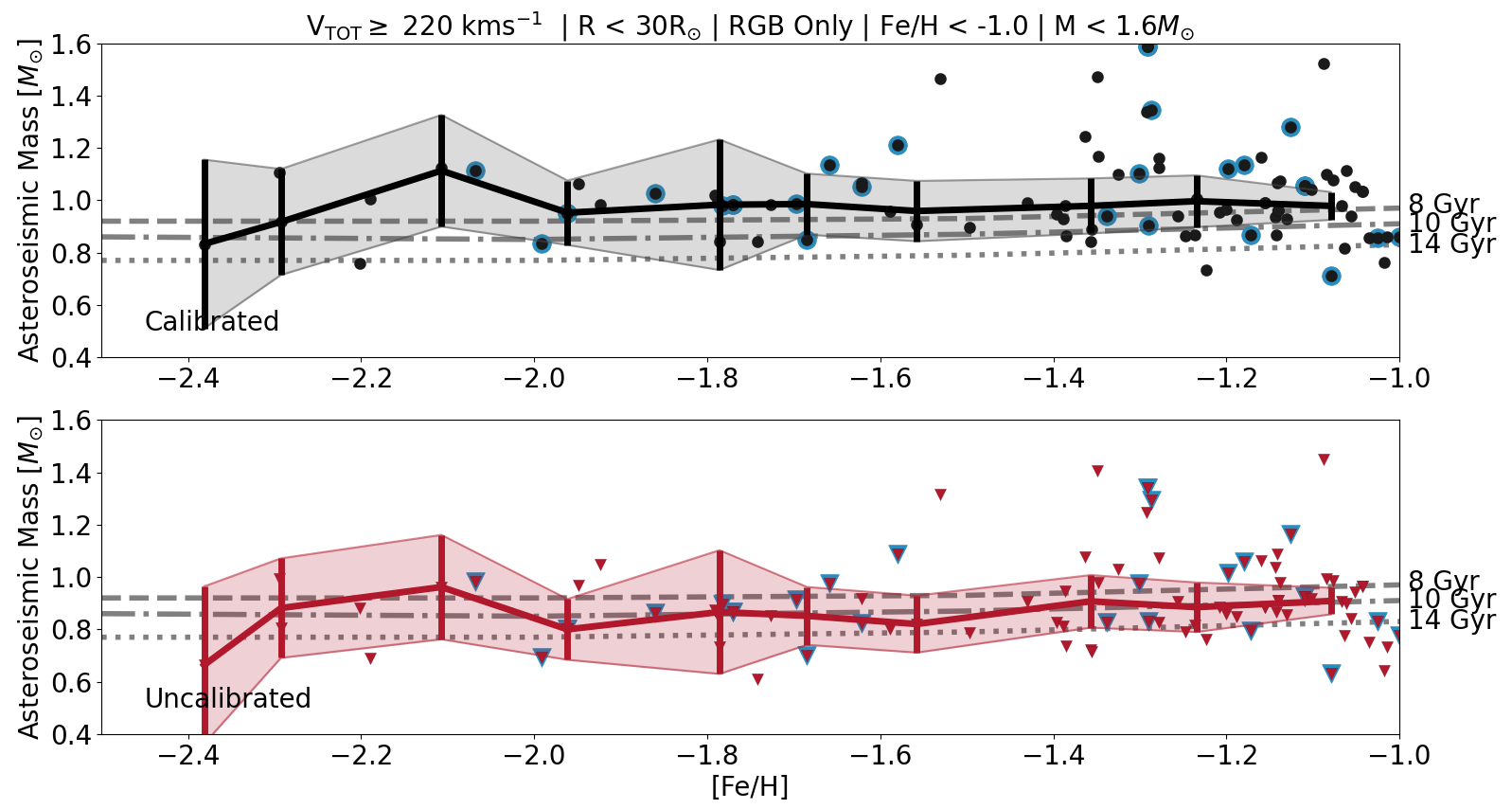}
\caption{Asteroseismic mass [M$_{\odot}$] as a function of APOGEE [Fe/H] [dex], calculated using asteroseismic scaling relations and mass and radius coefficients (as discussed in Section \ref{sec:sample_data_astero}). Dartmouth isochrones are shown by grey lines and labelled with their ages to the right of the plots (8 Gyr, 10 Gyr, and 14 Gyr). The black points in the top plot represent masses derived using calibrated temperatures from APOGEE and the existence of so many black points above the isochrone lines demonstrates the existence of potential overestimates in asteroseismic mass for the low-metallicity regime. Red triangles in the bottom plot represent masses calculated using the uncalibrated temperatures from APOGEE. The title shows cuts made to this plot (e.g., halo and RGB stars only).  Binned medians of both samples have been added using a black (red) line on the top (bottom) plot with 1$\sigma$ errors shown by the shaded regions. Any stars with the APOGEE flag `\texttt{STAR\_BAD}' have been removed and those with and ASPCAPFLAG $=$ 0.0 (corresponding to no warnings) are surrounding by blue. \label{fig:metal_mass}}
\end{figure*}

\begin{figure*}
\centering
\includegraphics[scale=0.75]{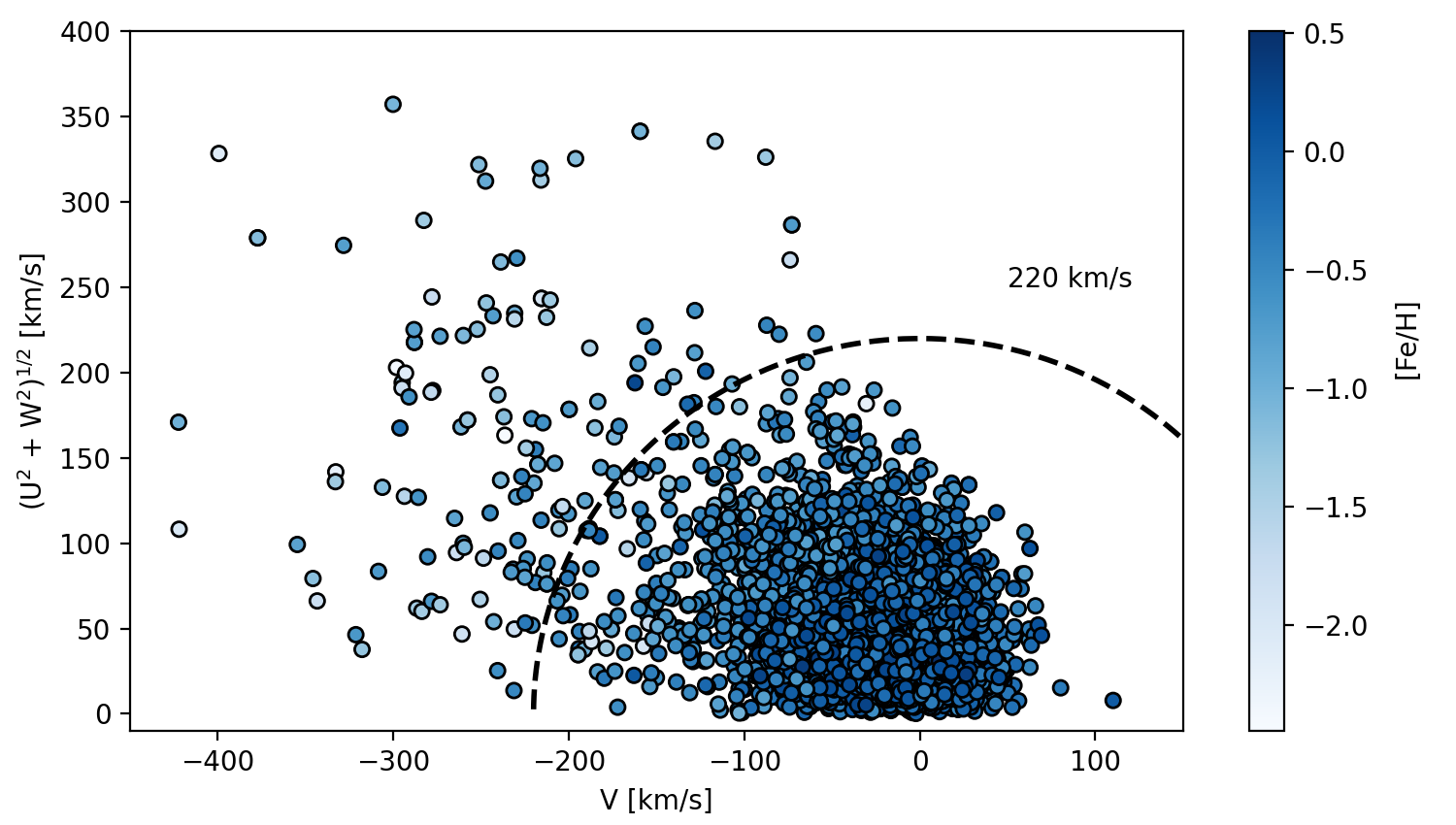}
\caption{A Toomre diagram of our sample, showing stellar velocity relative to the local standard of rest. The color-bar in this plot represents [Fe/H] from APOGEE and a dotted black line represents a velocity of 220 [km/s]. This line is used to delineate between halo stars and the rest of the sample, for use in Figure \label{fig:halo}.}
\end{figure*}

\citet{Epstein2014} showed that the asteroseismic masses determined through scaling relations in the APOKASC1 sample are higher than expected for the halo stars. Given that the halo stars are old, the median true mass in this regime should be lower. \citet{Sharma2016} applied the model-motivated $\Delta\nu$ corrections, f$_\Delta\nu$, to the scaling relations, which largely removed discrepancy reported by \citet{Epstein2014}. 

Other authors have investigated low-metallicity asteroseismic masses and compared them against \textit{Gaia}-derived masses \citep{Zinn2019a} and M4 cluster masses \citep{Miglio2016,Valentini2019,Matsuno2021,Tailo2022, Howell2022}. The latter studies, which also use the \citep{Sharma2016} $\Delta\nu$ corrections also find no evidence for the overmassive asteroseismic masses, in agreement with \citet{Sharma2016}. Another so far unconsidered aspect is a $\nu_{\mathrm{max}}$ metallicity dependence, as discussed by \cite{Viani2017}, who conclude that although the effects on main sequence and subgiant stars are minimal the effects on giant stars are much larger. With our relatively large sample of low metallicity stars, we therefore revisit the issue of seismic scaling-based mass for the low metallicity regime. 

In Figure \ref{fig:metal_mass} we show a selection of the low metallicity ($<$ -1.0 dex) halo stars (V $\geq$ 220 km$s^{-1}$) in our sample (See Figure \ref{fig:halo} for the graphic depiction of our halo star cut). To ensure that the stars with masses higher than expected are not the result of explainable factors we make the following cuts: all stars with the evolutionary type `RC' are removed as it is likely that the evolutionary states in this regime are wrong, because hot RC stars would not show oscillations. All stars with a mass $>$ 1.6 M$_{\odot}$ are removed to reduce the likelihood of merger products. There is also the recognition that there are asteroseismic biases in the highly luminous stars, so we remove stars with R $>$ 30R$_{\odot}$ \citep{Mosser2013,Stello2014,Kallinger2018,Zinn2019a}. 

We use Dartmouth isochrones \citep{Dotter2008}, with $\alpha$-enhancement (of $\alpha$ = 0.4), to demonstrate where we would expect our stars to fall\footnote{In Section \ref{sec:sub-populations} we use MIST models \citep{Dotter2016, Choi2016} instead of the Dartmouth isochrones used here. The use of Dartmouth at this point was to allow for an $\alpha$-enhancement factor that was not necessary for the AGB cut.}. We also removed stars with a `\texttt{STAR\_BAD}' flag. 

Initially, we plotted our stars using the calibrated temperatures from APOGEE (Top plot in Figure \ref{fig:metal_mass}). The calibrated spectroscopic temperatures available in APOGEE are calibrated to \cite{Gonzalez-Hernandez2009}. Full details are available in \cite{Holtzman2018}. Using the standard (f$_\Delta\nu$ = 1) scaling relation we find, similarly to \cite{Epstein2014}, that the asteroseismic masses of our metal-poor halo stars are much higher than Galactic models would suggest given the age of the Universe. 

Given the uncertain temperature calibration in the low-metallicity regime, we then considered the effect of using uncalibrated, ionization balance temperatures from APOGEE. Plotting the stars using the uncalibrated temperatures (Bottom plot in Figure \ref{fig:metal_mass}) we found much better agreement with the isochrones; this indicates that temperature calibration may be a key factor in the resulting mass values at low metallicities. The 2009 IRFM relationship \citep{Gonzalez-Hernandez2009} was anchored on a small sample of metal-poor stars, and a modest zero-point adjustment is reasonable. The mean mass of stars with the calibrated temperatures (black dots in Figure \ref{fig:metal_mass}) is 1.02 $\pm$ 0.25 M$_{\odot}$, and the mean mass using the uncalibrated temperatures (red triangles in Figure \ref{fig:metal_mass}) is 0.88 $\pm$ 0.24 M$_{\odot}$, indicating an overestimate of masses (using calibrated temperatures) of $\sim$14\% around the mean. This is similar to the value reported by \cite{Epstein2014}, describing an overestimate of 11\% $\pm$ 4\%. 

Two mass limits on this plot are of particular interest considering the precision of our asteroseismic masses. The first are the few RGB stars at M $<$ 0.8 M$_{\odot}$, which corresponds to the approximate minimum mass of an RGB star at the current age of the Galaxy (see mass for 14 Gyr isochrone with Dartmouth). These stars may have undergone mass loss throughout their evolution and represent interesting future studies. Conversely, 1.6 M$_{\odot}$ corresponds to the maximum possible mass (the maximum merger mass of two 0.8 M$_{\odot}$ stars). The RGB stars above this limit warrant further follow-up as they could represent interaction products.

\subsection{[$\alpha$/Fe]-Bimodality}
\label{sec:alpha_bimodality}

\begin{figure*}
\includegraphics[scale=0.70]{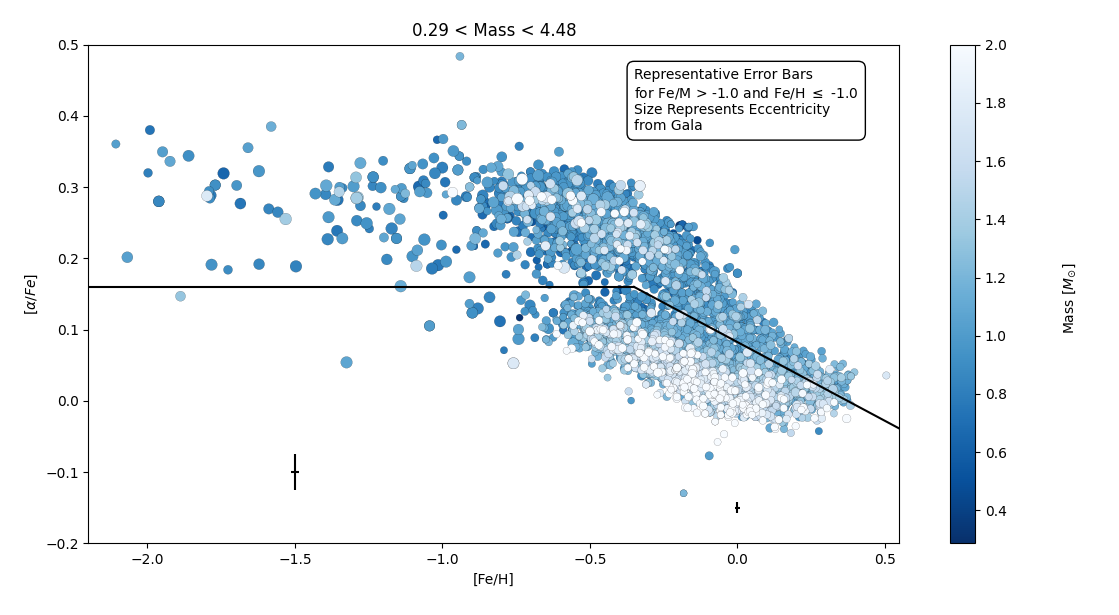}
\caption{[$\alpha$/Fe]-bimodality plot with the metallicity [Fe/H] ([$\alpha$/Fe]) on the x-axis (y-axis). Marker color represent the asteroseismic mass [M$_{\odot}$], which have been truncated for clarity, with the full range of masses show in the title. The darker blue colors correspond to lower masses, with the data sorted by mass so that higher mass stars appear on the top of the scatter. The size of each marker represents the Galactic eccentricity from \texttt{Gala}. A dividing line is drawn to separate the low- and high-[$\alpha$/Fe] stars, and this separation results in the flag contained in the APO-K2 catalog. Representative error bars are shown for both the relatively high and relatively low metallicity stars, where the separation occurs at [Fe/H] = -1.0. \label{fig:alpha_bi}}
\end{figure*}

\begin{figure}[ht!]
\includegraphics[scale=0.50]{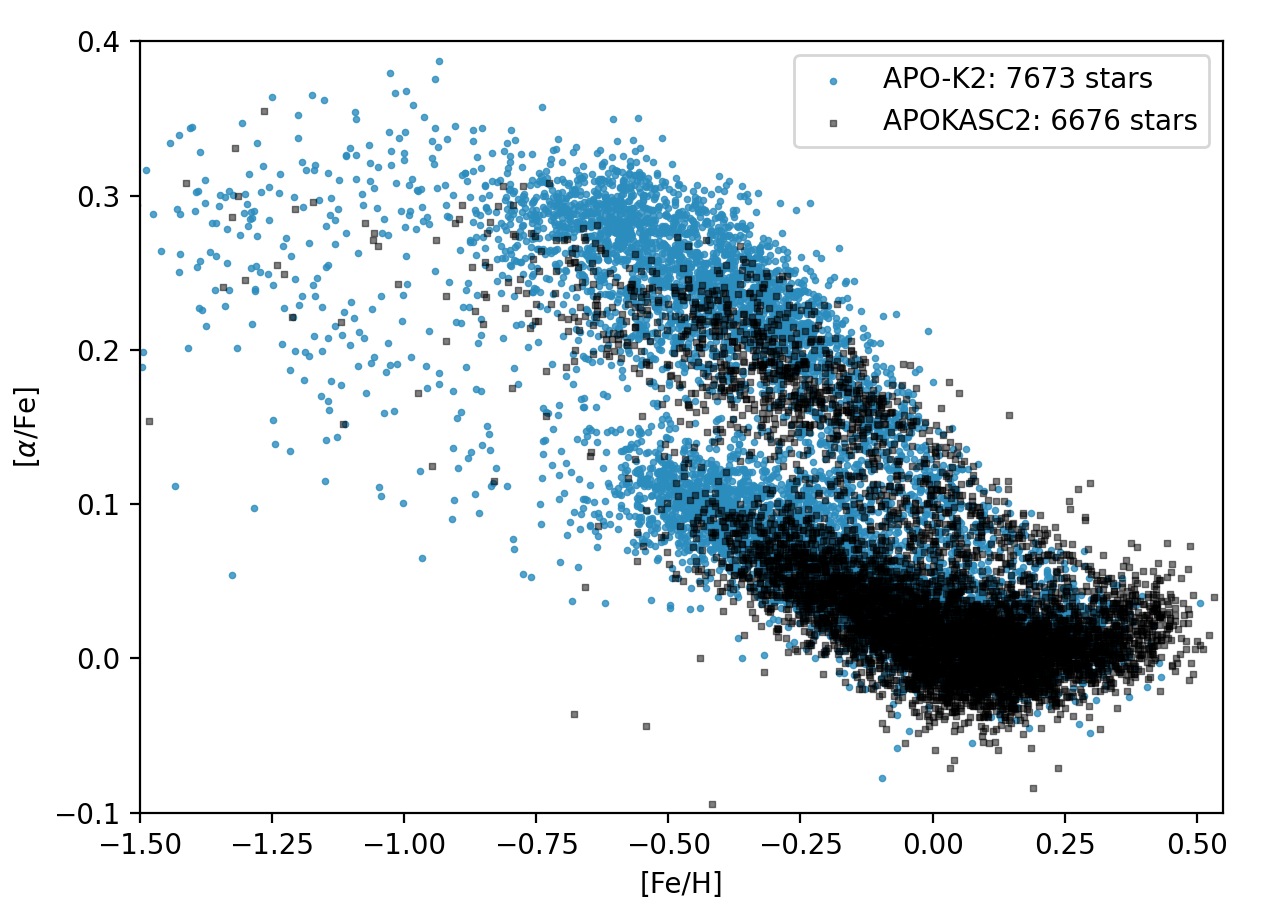}
\caption{$\alpha$-bimodality plot for both APOKASC2 (grey) and APO-K2 (blue) overplotted to highlight the increase of sample size in this parameter space.  \label{fig:alpha_bi_kepler}}
\end{figure}

An important abundance relation for exploring chemical enrichment in the Milky Way is the [$\alpha$/Fe]-bimodality. This association compares $\alpha$ process elements (e.g., O, Mg, Ca, and Si) to Fe abundance in stellar populations, which results in two groups, clearly separable on a plot of [Fe/H] vs. [$\alpha$/Fe], called the high- and low-[$\alpha$/Fe] stars. 

There are multiple theories about the origin of this double sequence. It has been suggested that the stars with larger $\alpha$ abundance form under different circumstances than the low-[$\alpha$/Fe] stars \citep{Mackereth2018}. An enhanced $\alpha$ with high Fe suggests that the majority of the heavy elements come from core-collapse SNe, whilst a low-[$\alpha$/Fe] mixture arises from a combination of SNe Ia and core-collapse SNe. The low-[$\alpha$/Fe] stars tend to be young, reside in the thin disk, and form their own sequence. One possibility is that they result from decreased star formation efficiency as the Galaxy ages \citep{Nidever2014}; they also show different birth radii and an anti-correlation between angular momentum and [Fe/H], which suggests the existence of radial migration could be needed to form the sequence \citep{Sharma2021}. The old, high-[$\alpha$/Fe] stars generally reside in the thick disk; they display enhanced $\alpha$ abundance, sometimes low metal abundance, and are relatively kinematically hot \citep[e.g.,][]{Haywood2013}. These stars can form from intense episodes of star formation, mainly from core-collapse supernova, or from gas polluted by core-collapse supernova. The [$\alpha$/Fe]-bimodality has been seen both in the solar neighbourhood and beyond \citep{Hayden2017}; recent studies using asteroseismic ages find the age distribution for high- and low-[$\alpha$/Fe] stars converge with increasing distance from the Galactic plane, support the [$\alpha$/Fe]-bimodality \citep{Warfield2021}, but questions still remain about whether the high-[$\alpha$/Fe] population is old.

Our K2 data allows investigation of the [$\alpha$/Fe]-bimodality. Figure \ref{fig:alpha_bi} shows the $\alpha$-bimodality plot for our sample. Asteroseismic masses are defined as in Section \ref{sec:sample_data_astero} and abundance information is taken from APOGEE. The eccentricity is defined in Section \ref{sec:kinematics}.  

Our data extends the abundance ranges to lower metallicities, higher $\alpha$ abundances, and farther distances than \textit{Kepler}. The APO-K2 sample offers a sample with asteroseismic masses including a known selection function to build on existing asteroseismic catalogs (e.g., \cite{Rendle2019,Mackereth2019,Imig2022}. 

Using the asteroseismic mass as a proxy for age, we can see a clump of higher mass stars in the low-[$\alpha$/Fe] regime, suggesting these stars are generally younger. They are also in a more condensed mass range than the high-[$\alpha$/Fe] stars, where the high-[$\alpha$/Fe] regime includes some stars with considerably lower mass, suggesting that they are older. We use the size of the marker to represent eccentricity and see that the highest eccentricity stars generally have high [$\alpha$/Fe]-abundance and low metallicity.

Our sample adds over 1,000 stars to the [$\alpha$/Fe]-bimodality plot as compared to the APOKASC2 sample ($\sim$6,000 stars), extending to lower metallicities and higher [$\alpha$/Fe] abundance, allowing us to more clearly separate the high- and low-[$\alpha$/Fe] samples. Figure \ref{fig:alpha_bi_kepler} shows both the APOKASC2 and APO-K2 catalogs over-plotted in metallicity-[$\alpha$/Fe], illustrating the extent to which the APO-K2 catalog has expanded on the \textit{Kepler} field, and confirming the convergence of the bimodality into a single distribution at higher metallicities.

\citet{Warfield2021} explored this space in K2 C4, C6, and C7, and discovered overlap between high- and low-$\alpha$ populations with stars of similar age. This discussion will continue in the companion paper of ages for this sample (Warfield et al. in prep.). 

\subsection{Kinematics}
\label{sec:kinematics}

\begin{figure*}
\includegraphics[scale=0.60]{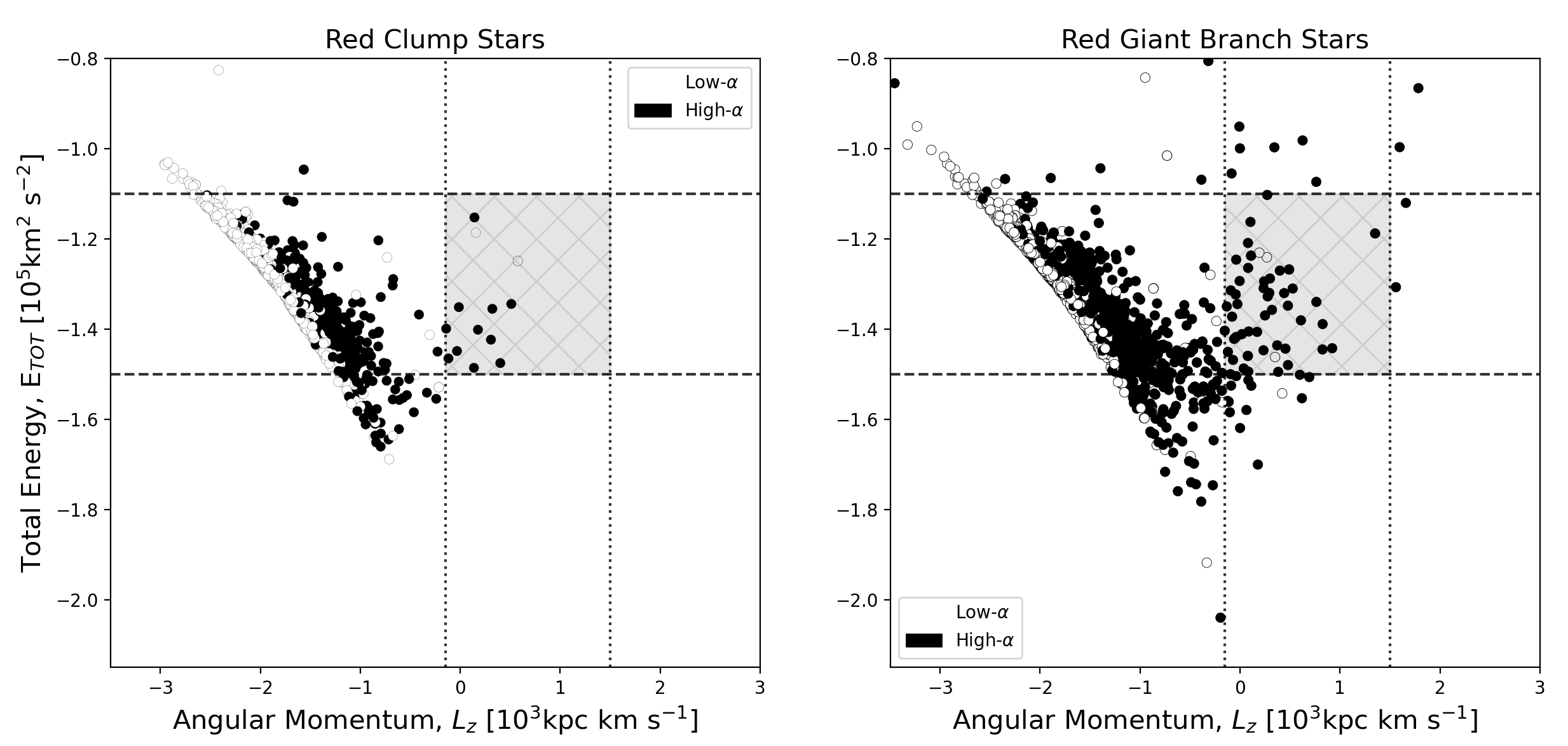}
\caption{Kinematic plots for the APO-K2 sample. The plot on the left (right) corresponds to RC stars (RGB stars). Only stars with positive \textit{Gaia} parallaxes are plotted. The grey area of the plots correspond to the area likely to host GES stars, and are defined by the lines drawn (and stated in the text). In both of these plots, the black (white) stars correspond to the high-[$\alpha$/Fe] (low-[$\alpha$/Fe]) stars, as defined by Figure \ref{fig:alpha_bi}.  \label{fig:energy_plot}}
\end{figure*}

The \textit{Gaia-Encaledus} (GES) structure is thought to represent the remnants of a dwarf galaxy that merged with the Milky Way \citep{Belokurov2018,Helmi2018,Montalban2021} in its early history. Though initially identified by kinematics, GES can also be identified by its combination of low metallicity and particular abundance pattern \citep{Haywood2018,Mackereth2019}. 

To define the dynamical information for our sample we use \texttt{Gala}, an \texttt{Astropy}-affiliated \citep{Astropy2013,Astropy2018} \texttt{Python} package. \texttt{Gala} uses the \texttt{Astropy} Galactocentric frame parameters adopted in \texttt{Astropy} v4.0. These are defined with a solar position of $R_{\odot}$ = 8.122 kpc and $z_{\odot}$ = 20.8 pc. The velocity of the Sun in the Galactocentric frame is (U, V, W)$_{\odot}$ = (12.9, 245.6, 7.78) km/s. For our analysis we adopt the Milky Way Potential available with \texttt{Gala} \citep{Bovy2015}, using the default parameters of Milky Way mass, virial radius etc. In Section \ref{sec:alpha_bimodality} we used eccentricities ($e$) from \texttt{Gala} defined as

\begin{equation}
    e = \frac{r_{apo} - r_{peri}}{r_{apo} + r_{peri}},
\end{equation}
where $r_{apo}$ is the orbital apocentre and $r_{peri}$ is the orbital pericentre. 

Figure \ref{fig:energy_plot} shows orbital angular momentum (L$_z$) as a function of orbital energy (E). This space is most often used to identify merger debris from past accretion events. Colors correspond to the high-[$\alpha$/Fe] (green) and low-[$\alpha$/Fe] (purple) samples. We display the sample broken down by evolutionary state (RGB and RC), with boundaries for the GES overlaid. The vertical lines denote the GES limits in angular momentum from \citet{Helmi2018}, between $-$150 $<$ L$_z$ [kms$^{-1}$kpc] $<$ 1500. The GES distinction in energy is taken from \citet{Koppelman2019} and placed between $-$1.1$\times10^{5}$ km$^{2}$s$^{-2}$ and $-$1.5$\times10^{5}$ km$^{2}$s$^{-2}$. Inside the grey box lie a few dozen GES substructure candidates. 
 
\citet{Koppelman2019} discuss the necessity of a chemical tagging analysis to determine whether substructures are related to accretion events. They studied the distribution of nearby thick disk and halo stars using 6D phase-space data from \textit{Gaia} DR2 and found that not all substructure is due to accretion, nor is it due to the settling of the gravitational potential after major activity \citep{Haywood2018}. The range of kinematic parameter space probed in this sample may prove useful for future analysis, in combination with the selection functions presented in S22.

Figure \ref{fig:energy_plot}, shows the low-[$\alpha$/Fe] stars mainly occupy the disk, and hence are seen in fairly circular orbits that sit close to the hyperbola that defines the minimum energy curve in this space. By contrast, the high-[$\alpha$/Fe] stars, residing mainly in the thick disk and halo are expected to be kinematically hot (possessing eccentric orbits) and also occupy regions above the minimum energy hyperbola, but with a larger spread.

\section{Conclusions}
\label{sec:conclusion}

In this paper we summarize the APO-K2 evolved star sample, with corresponding spectroscopic (APOGEE DR17), asteroseismic (K2-GAP), and astrometric (\textit{Gaia} EDR3) parameters. Our sample of 7,673 unique stars contains RGB, RC, secondary red clump, and RGB bump evolutionary states, from various areas of the Galaxy. Our work provides precise asteroseismic radii and masses as well as evolutionary states and metallicities, explored in multiple parameter spaces. 

Throughout this work, we overviewed many parameter spaces that our catalog extends, some for the first time. We investigate the completeness of our sample by comparing it to the S22 selection function, in the mass-radius, mass-metallicity, color-magnitude, and magnitude-$\nu_{\mathrm{max}}$ regimes. We examine our astereosismic masses in the low-metallicity regime, resulting in higher masses than expected for the low-metallicity stars, even when taking corrections to the $\Delta\nu$ sclaing relation into account. We find that using raw APOGEE temperatures to derive stellar masses results in a better agreement with astrophysical estimates for very metal-poor stars. As specific science applications, we also identify possible stellar merger products. Further, we use this low-metallicity sample to extend our understanding of the $\alpha$-bimodality in new regions of parameter space. Finally, we looked at our sample in energy/kinematic space, with \textit{Gaia} EDR3, and identified potential GES stars.

The overview presented in our paper only scratches the surface of the rich data sample; some of the spaces explored in this paper will be further investigated in follow-up papers. Further work will be undertaken to explore the multiplicity (Schonhut-Stasik et al. in prep.), and abundance space provided by APOGEE (Schonhut-Stasik et al. in prep.), for example, to investigate the carbon-enhanced stars known to exist at low metallicity ([Fe/H] $<$ 2.0 in \cite{Beers2005} and [Fe/H] $\leq$ -2.5 [dex] in \citet{Suda2008}). Furthermore, a deeper analysis of interesting individual objects, such as high-mass stars at low metallicities and potential GE/merger candidates will be explored. A companion paper will release age information (Warfield et al. in prep.). 

The amount of stars accessible to Galactic archaeology using asteroseismology will only grow with future missions. For example, the NASA planet-finding mission TESS \citep{Ricker2014, Hon2022} and the Nancy Grace Roman Telescope \citep{Gould2015, Spergel2015}, as well as the ESA missions Euclid \citep{Laureijs2011, Gould2016} and PLATO \citep{Rauer2014, Miglio2017}, will yield enormous harvests of asteroseismic detections. In terms of spectroscopic measurements, the upcoming projects WEAVE \citep{Dalton2020}, MOONS \citep{Cirasuolo2014}, and 4MOST \citep{DeJong2012} will increase chemical abundance yields. Finally with future releases from \textit{Gaia}, our astronometry and kinematic data will only increase in precision.

\begin{table*}[t]
\centering
\begin{tabular}{|c|c|c|c|c|c|c|c|c|c|}
\hline
EPIC & APOGEE ID & Campaign & Gal Lat & Gal Long & Evol & $T_{\mathrm{eff}}$ & $T_{\mathrm{eff}}$ & log(g) & log(g) \\
& & & [Deg] & [Deg] & State & [K] & Error [K] & [dex] & Error [dex]  \\
\hline
\hline
211705076 & 2M08374542+1558546 & 5,16,18 & 30.556 & 209.677 & RGB & 4664 & 7.37 & 2.87 & 0.021 \\ 
212873074 & 2M13305142-0302596 & 17 & 58.359 & 321.945 & RGB & 4789 & 11.35 & 2.78 & 0.029 \\ 
204651310 & 2M16593864-2127129 & 11 & 12.794 & 0.400 & RGB & 4667 & 7.05 & 3.03 & 0.020 \\ 
212386819 & 2M13364485-1513074 & 18,16 & 46.249 & 318.838 & RC & 4761 & 8.04 & 2.47 & 0.022 \\ 
228774437 & 2M12441419-0751315 & 10 & 54.969 & 299.824 & RC & 5006 & 9.48 & 2.85 & 0.021 \\ 
212349887 & 2M13394793-1608035 & 1 & 45.174 & 319.515 & RC & 4828 & 8.84 & 2.46 & 0.024 \\ 
204970434 & 2M16494197-2001358 & 12 & 15.510 & 0.145 & RGB & 4827 & 13.78 & 2.42 & 0.035 \\ 
228909546 & 2M12254205-0306337 & 3 & 59.131 & 290.334 & RC & 4725 & 7.78 & 2.38 & 0.0217 \\ 
...&...&...&...&...&...&...&...&...&... \\
\hline
\end{tabular}

\vspace{0.5cm}

\begin{tabular}{|c|c|c|c|c|c|c|c|c|c|}
\hline
[M/H] & [Fe/H] & $[\alpha/M]$ & $\alpha$ flags & $V_{\mathrm{mag}}$ & J-K & $\nu_{max}$ & $\Delta\nu$ & $f\Delta\nu$ & $\kappa_{M}$ \\
& & & & & & $\mu$Hz & $\mu$Hz & & [$M_{\odot}$] \\
\hline
\hline
0.00 & -0.01 & 0.07 & 1 & 9.25 & 0.64 & 98.0 & 9.27 & 1.024 & 1.463 \\ 
-0.27 & -0.28 & 0.06 & 1 & 6.87 & 0.66 & 49.9 & 5.20 & 1.026 & 1.955 \\ 
0.19 & 0.17 & 0.03 & 1 & 9.60 & 0.734 & 148.6 & 12.95 & 1.016 & 1.338 \\ 
-0.05 & -0.06 & 0.01 & 1 & 9.34 & 0.64 & 33.0 & 3.95 & 0.998 & 1.686 \\ 
0.04 & 0.03 & -0.02 & 1 & 9.05 & 0.57 & 89.2 & 7.56 & 0.993 & 2.495 \\ 
-0.24 & -0.25 & 0.05 & 1 & 9.59 & 0.69 & 31.6 & 3.97 & 0.998 & 1.457 \\ 
-0.45 & -0.46 & 0.27 & 2 & 13.19 & 0.70 & 26.0 & 2.98 & 1.021 & 2.554 \\ 
-0.03 & -0.03 & -0.00 & 1 & 9.35 & 0.68 & 31.1 & 3.69 & 0.997 & 1.857 \\ 
...&...&...&...&...&...&...&...&...&... \\
\hline
\end{tabular}

\vspace{0.5cm}

\begin{tabular}{|c|c|c|c|c|c|c|c|c|}
\hline
Mass & Mass & $\kappa_R$ & Radius & Radius & Galactic & Angular Momentum & Total Energy & $f\Delta\nu$ Flag \\
M$_{\odot}$ & Error [$M_{\odot}$] & [$R_{\odot}$] & [$R_{\odot}$] & Error [$R_{\odot}$] & Eccentricity & [kpc km s$^{-1}$] & [km$^{2}$ s$^{-2}$] & \\
\hline 
\hline 
1.06 & 0.069 & 6.776 & 6.09 & 0.125 & 0.84 & -6.5 & -2711.8 & 0 \\ 
1.48 & 0.050 & 10.974 & 9.99 & 0.111 & 0.90 & -58.3 & -2890.9 & 0 \\ 
0.97 & 0.033 & 5.262 & 4.73 & 0.053 & 0.99 & -346.9 & -2783.3 & 0 \\ 
1.26 & 0.196 & 12.540 & 11.38 & 0.575 & 0.97 & 253.4 & -2594.4 & 0 \\ 
2.01 & 0.189 & 9.273 & 8.63 & 0.261 & 0.95 & -1662.3 & -2632.9 & 0 \\ 
1.11 & 0.097 & 11.908 & 10.89 & 0.309 & 0.60 & -3263.7 & -2551.1 & 0 \\ 
1.95 & 0.217 & 17.386 & 15.89 & 0.576 & 0.85 & -2641.1 & -1960.7 & 0 \\ 
1.37 & 0.370 & 13.557 & 12.26 & 1.048 & 0.79 & -1210.8 & -2580.7 & 0 \\ 
...&...&...&...&...&...&...&...&... \\
\hline
\end{tabular}

\caption{\label{tab:table_description}APO-K2 catalog. EPIC column gives the EPIC ID number for the star, and corresponds to the same row in each of the smaller tables. All observed campaigns are listed in the second column, followed by the Galactic Latitude and Longitude [degrees]. Next, we show the evolutionary state, as defined in the text as either red clump (RC) or red giant branch (RGB). The remaining columns in the top section of the table represent T$_{\mathrm{eff}}$ [K], and log(g) [dex] with their associated errors. The second section of the table gives spectroscopic abundance values from APOGEE: [M/H], [Fe/H], and [$\alpha$/M]. Next the alpha flag is given, with 1 corresponding to high [$\alpha$/Fe], 0 to low [$\alpha$/Fe], and -1 to the middle of the [$\alpha$/Fe] bimodality, with a 95\% confidence interval around the dividing line. Next, we provide the V-band magnitude and the j-k colors used in the selection function plots The final four columns of this section are the frequency of maximum oscillation power [$\mu Hz$] and large frequency separation [$\mu Hz$], followed by the f$\Delta\nu$ value. The final column of this section of the table shows the coefficient of mass [M$_{\odot}$]. The stars mass, with the coefficient used, and the appropriate error come first in the bottom section of the table, followed by the radius coefficient, stellar radius, and radius error (all [R$_{\odot}$]). Finally we provide the kinematic information from \texttt{Gala} with values of Galactic eccentricity, angular momentum and total energy. The final column is the f$\Delta\nu$ flag defined in the text. Full version available electronically, in one table.}

\end{table*}

\section*{Acknowledgements}

We would like to thank the community that supports APO-K2. J.S.S. is supported a Neurodiversity Inspired Science and Engineering (NISE) Fellowship under award. J.C.Z. is supported by an NSF Astronomy and Astrophysics Postdoctoral Fellowship under award AST-2001869. D.M. gratefully acknowledges support by the ANID BASAL projects ACE210002 and FB210003 and by Fondecyt Project No. 1220724. S.M. acknowledges support from the Spanish Ministry of Science and Innovation (MICINN) with the Ram\'on y Cajal fellowship no. $\sim$RYC-2015-17697, grant no. PID2019-107187GB-I00 and PID2019-107061GB-C66 for PLATO, and through AEI under the Severo Ochoa Centres of Excellence Programme 2020--2023 (CEX2019-000920-S). D.S. acknowledges support from the Australian Research Council Discover Project  DP190100666. R.A.G. and B.M. acknowledge the support from the PLATO Centre National D'{\'{E}}tudes Spatiales grant.

This paper includes data collected by the \textit{Kepler} mission and the K2 mission. Funding for the \textit{Kepler} mission and K2 mission are provided by the NASA Science Mission directorate. 

This work has made use of data from the European Space Agency (ESA) mission \emph{Gaia} (\url{https://www.cosmos.esa.int/gaia}), processed by the\emph{Gaia} Data Processing and Analysis Consortium (DPAC, \url{https://www.cosmos.esa.int/web/gaia/dpac/consortium}). Funding for the DPAC has been provided by national institutions, in particular the institutions participating in the \emph{Gaia} Multilateral Agreement.
This research has made use of the VizieR  catalogue access tool, CDS, Strasbourg, France (DOI: 10.26093/cds/vizier). The original description of the VizieR  service was published \citet{vizier2000}.

Funding for the Sloan Digital Sky Survey IV has been provided by the Alfred P.~Sloan Foundation, the U.S. Department of Energy Office of Science, and the Participating Institutions. SDSS-IV acknowledges support and resources from the Center for High-Performance Computing at the University of Utah. The SDSS web site is \url{www.sdss.org}.

SDSS-IV is managed by the Astrophysical Research Consortium for the Participating Institutions of the SDSS Collaboration including the Brazilian Participation Group, the Carnegie Institution for Science, Carnegie Mellon University, the Chilean Participation Group, the French Participation Group, Harvard-Smithsonian Center for Astrophysics, Instituto de Astrof\'isica de Canarias, The Johns Hopkins University, Kavli Institute for the Physics and Mathematics of the Universe (IPMU) / University of Tokyo, Lawrence Berkeley National Laboratory, Leibniz Institut f\"ur Astrophysik Potsdam (AIP),  Max-Planck-Institut f\"ur Astronomie (MPIA Heidelberg), Max-Planck-Institut f\"ur Astrophysik (MPA Garching), Max-Planck-Institut f\"ur Extraterrestrische Physik (MPE), National Astronomical Observatories of China, New Mexico State University, New York University, University of Notre Dame, Observat\'ario Nacional / MCTI, The Ohio State University, Pennsylvania State University, Shanghai Astronomical Observatory, United Kingdom Participation Group, Universidad Nacional Aut\'onoma de M\'exico, University of Arizona, University of Colorado Boulder, University of Oxford, University of Portsmouth, University of Utah, University of Virginia, University of Washington, University of Wisconsin, Vanderbilt University, and Yale University.
Land Acknowledgement?
This paper includes data collected by the Kepler mission and obtained from the MAST data archive at the Space Telescope Science Institute (STScI). Funding for the Kepler mission is provided by the NASA Science Mission Directorate. STScI is operated by the Association of Universities for Research in Astronomy, Inc., under NASA contract NAS 5–26555.

\software{Python 3 \citep{VanRossum2009}, numpy \citep{Harris2020}, matplotlib \cite{Hunter2007}, pandas \citep{Mckinney2010}, Astropy \citep{Astropy2013,Astropy2018}, mw\_plot ({\url{https://milkyway-plot.readthedocs.io/en/latest/}})}

\facilities{Du Pont (APOGEE), 
            Sloan (APOGEE), 
            2MASS,
            Kepler,
            K2}

\bibliography{main}{}
\bibliographystyle{aasjournal}



\end{document}